\documentclass[authoryear,3p,twocolumn]{elsarticle}

\usepackage{graphicx}
\usepackage{subfig}
\usepackage{wasysym}
\usepackage{ltxtable}
\usepackage{tabularx}
\usepackage{longtable}
\usepackage{lipsum}
\usepackage{lscape}
\usepackage{color}
\usepackage{epsf}
\usepackage{hyperref}

\begin{document}
\title{Chandra Observations of the HII Complex G5.89-0.39 and TeV Gamma-Ray Source HESSJ1800-240B}
\author[adel]{E.J.~Hampton\corref{cor1}}
\ead{elise.hampton@anu.edu.au}
\author[adel]{G.~Rowell}
\author[mp]{W.~Hofmann}
\author[EP]{D.~Horns}
\author[RU]{Y.~Uchiyama}
\author[UH]{S.~Wagner}
\cortext[cor1]{Current institution: Research School of Astronomy \& Astrophysics, Australian National University, Mt Stromlo Observatory, Cotter Road, Weston 2611, ACT, Australia} 
\address[adel]{School of Chemistry \& Physics, University of Adelaide, Adelaide 5005, Australia}
\address[mp]{Max Planck Instit\"ut f\"ur Kernhysik, P.O. Box 103980, D 69029 Heidelberg, Germany}
\address[EP]{Institut fuer Experimental Physik, Universitaet Hamburg, Luruper Chausse 149, D22761 Hamburg Germany}
\address[RU]{Department of Physics, Rikkyo University, 3-34-1 Nishi-Ikebukuro, Toshima-ku, Tokyo, Japan 171-8501}
\address[UH]{Landessternwarte, Universitaet Heidelberg, Koenigtuhl, D69117 Heidelberg, Germany}

\begin{abstract}
We present the results of our investigation, using a Chandra X-ray observation, into the stellar population of the massive star formation 
region G5.89-0.39,  and its potential connection to the 
coincident TeV gamma-ray source HESSJ1800-240B. G5.89-0.39 comprises two separate HII regions G5.89-0.39A and G5.89-0.39B (an ultra-compact HII region). 
We identified 159 individual X-ray point sources in our observation using the source detection algorithm \texttt{wavdetect}. 
35 X-ray sources are associated with the HII complex G5.89-0.39. The 35 X-ray sources represent an average unabsorbed luminosity (0.3-10\,keV) of $\sim10^{30.5}$\,erg/s,
typical of B7-B5 type stars. The potential ionising source of G5.89-0.39B known as Feldt's star is possibly identified in our observation with an unabsorbed X-ray luminosity suggestive of a B7-B5 star.
The stacked energy spectra of these sources is well-fitted with a single thermal plasma APEC model with kT$\sim$5\,keV, and column density N$_{\rm H}=2.6\times10^{22}$\,cm$^{-2}$ (A$_{\rm V}\sim 10$). The residual (source-subtracted) X-ray emission towards G5.89-0.39A and B 
is about 30\% and 25\% larger than their respective stacked source luminosities. Assuming this residual emission is from unresolved stellar sources, the total B-type-equivalent stellar 
content in G5.89-0.39A and B would be 75 stars, consistent with an earlier estimate of the total stellar mass of hot stars in G5.89-0.39.
We have also looked at the variability of the 35 X-ray sources in G5.89-0.39. Ten of these sources are flagged as being variable. Further studies are needed to determine the exact causes of the variability, however the variability could point towards pre-main sequence stars.
Such a stellar population could provide sufficient kinetic energy to account for a part of the GeV to TeV gamma-ray emission in the source HESSJ1800-240B. However, future arc-minute
angular resolution gamma-ray imaging will be needed to disentangle the potential gamma-ray components powered by G5.89-0.39 from those powered by the W28 SNR.
\end{abstract}

\maketitle

\section{Introduction}
\label{sec:intro}
 The massive star formation region (SFR) and HII complex G5.89-0.39 is one of the best examples of a growing subset of galactic TeV ($10^{12}$ eV) gamma-ray sources coincident with the molecular gas associated with SFRs \citep[e.g.][]{Aharonian:2007aa,Chaves:2008aa,Ohm:2010,Eger:2011aa,MAGIC:2014aa,MAGIC:2014ab}. G5.89-0.39 is comprised of two star forming regions and is located $\sim 1^\circ$ south of the mature ($>10^4$\,yr old) mixed-morphology supernova remnant (SNR) W28. 

The TeV gamma-ray emission, peaking in the vicinity of the densest molecular gas of G5.89-0.39, indicates the presence of non-thermal multi-TeV protons (i.e. cosmic-rays) and/or electrons. The close spatial match between the GeV to TeV gamma-ray emission and the molecular gas of G5.89-0.39 plus the continuous spectrum from 100\,MeV to about 10\,TeV \citep{Aharonian:2008aa,Nolan:2012,Hanabata:2014aa} implies a cosmic-ray origin for the gamma-rays. The favoured scenario is that cosmic-rays accelerated and confined within the W28 SNR bubble eventually escape to collide with the surrounding gas. The GeV to TeV gamma-ray spectral shapes observed in HESSJ1800-240B and adjacent TeV sources such as HESSJ1800-240A and HESSJ1801-233 seem to fit well with the energy dependent diffusion of the particles to the gas clouds \citep[e.g.][]{Gabici:2010,Fujita:2009aa,Li:2010aa,Nava:2013aa} from W28. 

SFRs themselves have also been suggested as sites of particle acceleration. The shocks generated by proto-stellar jets interacting with the interstellar medium (ISM), the winds of massive stars, wind-wind collisions in binary systems, and the collective effects of winds in clusters have long-been put forward as potential ways to accelerate particles from GeV to multi-TeV energies \citep[see e.g.][]{Voelk:1982,Cesarsky:1983,Eichler:1993,Domingo:2006,Bednarek:2007,Reimer:2007,Araudo:2007aa,Bosch-Ramon:2010aa,Araudo:2014aa}. Some of these ideas have been motivated by the non-thermal radio emission (tracing GeV electrons) associated with the proto-stellar/ISM interaction region in several systems, but also by the obvious fact that massive stars provide a significant amount of stellar wind kinetic energy. 

Prior X-ray studies have revealed high plasma temperatures and/or high energy particle acceleration confirming the role of stellar winds and outflows as significant sources of energy. \citet{Townsley:2003aa} and \cite{Oskinova:2010aa} summarise a number of massive SFRs exhibiting diffuse X-ray emission. \citet{anderson11} attributed hard thermal X-ray emission to the collision of stellar winds in binary stars or star clusters. In such cases the shocks created by the colliding stellar winds can heat thermal plasma leading to high temperatures \citep[e.g.][]{Ozernoy:1997aa,Canto:2000}. The soft diffuse X-ray emission observed from the HII regions W49A $\rm{G_{X}}$ \citep{Tsujimoto:2006aa}, M17 and the Rosette Nebula  \citep{Townsley:2003aa} have been interpreted using the so-called wind blown bubble model \citep{Townsley:2003aa,Chu:2003aa,Wrigge:2005aa}. For M17 and the Rosette Nebula, their soft diffuse luminosities, which are driven by several O-type stars are between 10 to 50\% of their total point source X-ray luminosities \citep{Townsley:2003aa}. Wind-blown bubbles form due to strong stellar winds from massive stars colliding with the surrounding ionised gas in the HII region \citep{Capriotti:2001aa}, which then subsequently create a shock front. \citet{Pravdo:2009aa} summarise a number of protostellar outflows exhibiting thermal X-ray emission. Evidence for non-thermal X-ray emission has so far been seen in three massive stellar clusters \citep{Muno:2006,Oskinova:2010aa,Krivonos:2013aa} and in a protostellar/jet ISM interaction region \citep{Lopez-Santiago:2013aa}.

G5.89-0.39 harbours two distinct HII regions, G5.89-0.39A and B, with the B component known to be an ultra-compact HII region. Distance estimates for G5.89-0.39, from studies on the ionising source of G5.89-0.39B, Feldt's star \citep[][Section \ref{sec:feldt}]{Feldt:2003vn}, appear to be consistent with that of W28, at $d\sim$2\,kpc, although maser parallax measurements have suggested differing values ($d\sim1.28^{+0.09}_{-0.08}$\,kpc and $d\sim2.99^{+0.19}_{-0.17}$\,kpc from \citet{Motogi:2011} and \citet{Sato:2014}, respectively). 

The source of ionisation in G5.89-0.39B is believed to be a single star - Feldt's star \citep{Feldt:2003vn}, although a multiple stellar system has been discussed. Feldt's star has a temperature $\rm{\sim 40,000K}$ and mass $\rm{ > 200 M_{\astrosun}}$, indicating a possible O5 or earlier spectral type \citep{Feldt:2003vn,Puga:2006aa}. G5.89-0.39 has been intensely studied, with much of the focus on the energetic outflows associated with G5.89-0.39B \citep[e.g.][]{Harvey:1988aa,Churchwell:1990aa,Acord:1997aa,Kim:2003aa,Sollins:2004aa}. The outflows plus the suspected clusters of massive stars in the vicinity may provide an additional high energy particle component to that of the nearby W28 SNR in generating the gamma-ray emission of this region (discussed in section~\ref{sec:extended}).

To ascertain this possibility, it is essential to understand the stellar content of G5.89-0.39. As is typical for HII regions, dense molecular gas can hamper a detailed census of the stellar content via radio, infrared and optical observations. We have therefore turned to X-ray observations at arc-second resolution with the Chandra Observatory to carry out this task.

\begin{figure*}
  \centering
  \includegraphics[width=0.7\textwidth]{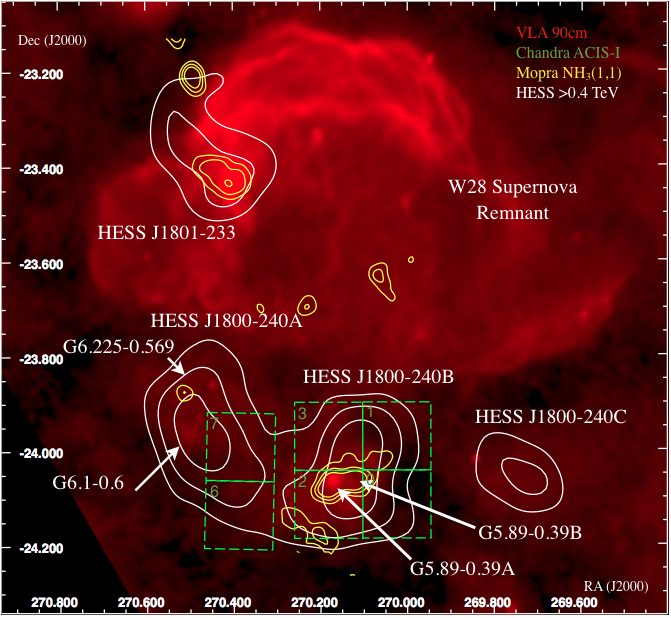}\\ 
  \includegraphics[width=0.7\textwidth]{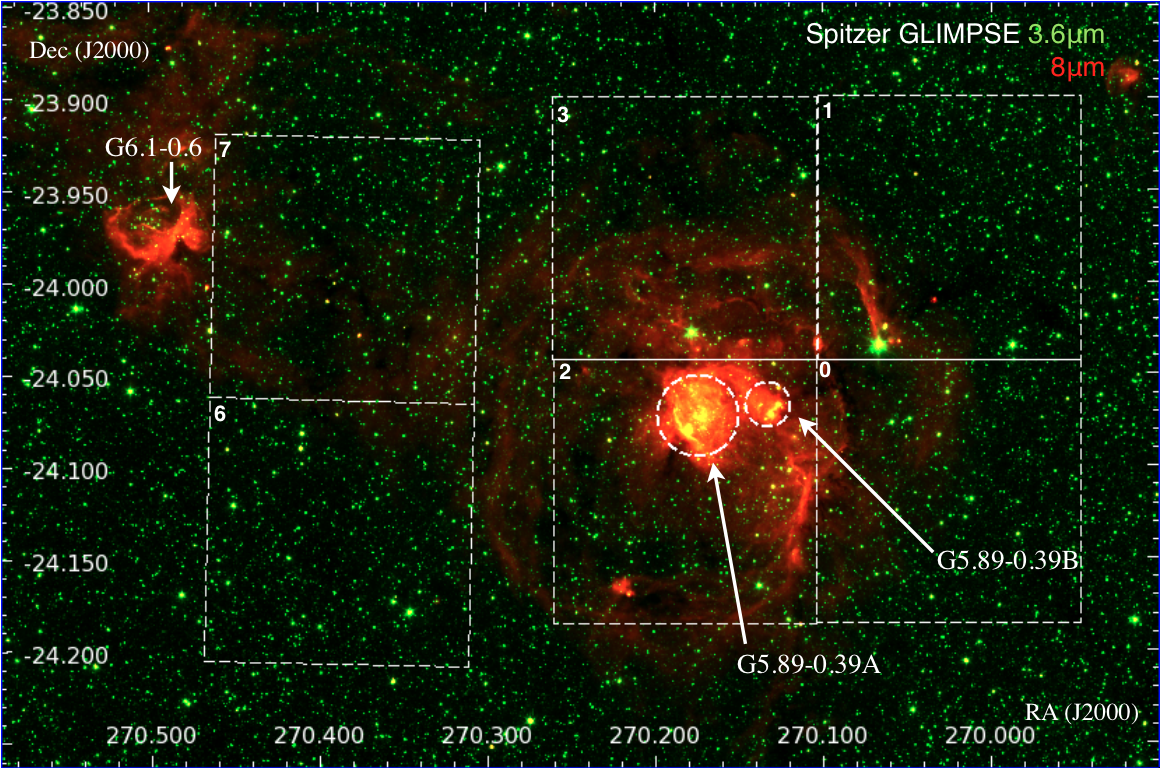}\\ 
  \caption{Top - 90cm VLA Radio observation of W28 (red), Mopra $\rm{NH_3(1,1)}$ dense gas observation (velocity integrated between 20-50 km/s -- yellow contours), H.E.S.S $\textgreater$ 0.4 TeV (white contours) 
    with TeV sources labelled. Green dashed lines indicate the field of view (FoV) of the Chandra ACIS-I CCDs, labelled by ID number. The HII complex G5.89-0.39 towards HESSJ1800-240B, is positioned 
    in CCD 2 with the conventional HII region, G5.89-0.39A, and the UC HII region, G5.89-0.39B, shown by the white arrows. Bottom - Zoom-in of the Spitzer $\rm{8\mu m}$ emission showing the ionised gas surrounding G5.89-0.39. The stellar content is highlighted by the $\rm{3.6\mu m}$ (green) emission in the HII complex as yellow points. White dashed 
    lines indicate the FoV of the Chandra ACIS-I CCDs. Also indicated are locations of the additional HII regions 6.225-0.569 and G6.1-0.6 towards HESSJ1800-240A.}
  \label{fig:main}
\end{figure*}

Figure~\ref{fig:main} (top panel) presents the VLA 90cm \citep{Brogan:2006aa} image with contour overlays from the Mopra radio telescope of $\rm{NH_{3}(1,1)}$ \citep{Nicholas:2011aa} and H.E.S.S TeV gamma-ray \citep{Aharonian:2008aa} observations of the W28 region including G5.89-0.39. The bright radio shell of W28 is clearly seen in the VLA 90cm observation. The Mopra $\rm{NH_{3}(1,1)}$ observations reveal the location of dense ($>10^4$\,m$^{-3}$) molecular gas, only the strongest $\rm{NH_{3}(1,1)}$ contours are shown in Figure~\ref{fig:main}. Figure~\ref{fig:ch3oh-cs10} (in appendix) shows the Mopra CS(1-0) dense gas tracer and the CH$_3$OH-I maser emission tracing ongoing star formation in the field of view of our Chandra observation \citep[from][]{Nicholas:2012aa} for comparison with Figure~\ref{fig:main}. Earlier studies \citep{Aharonian:2008aa} showed that all of the TeV gamma-ray sources are spatially coincident with CO(1-0) emission, tracing the moderately dense ($10^{1-2}$\,m$^{-3}$) diffuse molecular gas.   

One of the two brightest such gamma-ray sources, HESSJ1800-240B, clearly overlaps G5.89-0.39. Despite the proximity of G5.89-0.39 to W28, 14\,pc ($\sim0.4^{\circ}$) south of the edge of W28, the W28 shock is not suspected to be influencing G5.89-0.39 as revealed by the molecular gas dynamics studies of \citet{Nicholas:2011aa,Nicholas:2012aa}. The other bright source to the north east, HESSJ1801-233, is associated with the shocked molecular cloud region of W28 whilst HESSJ1800-240A to the west encompasses two other HII regions G6.1-0.6 and 6.225-0.569 which are also likely associated with same molecular cloud complex harbouring G5.89-0.39 \citep{Nicholas:2011aa}.

The bottom panel of Figure \ref{fig:main} presents an image of the Spitzer GLIMPSE $\rm{8\mu m}$ (red) and $\rm{3.6\mu m}$ (green) observations towards G5.89-0.39. The FoV of our Chandra observations are indicated by the dashed white boxes and labelled according to their Chandra ACIS-I CCD ID numbers. Based on the $8\mu m$ observations, we have estimated the diameters of G5.89-0.39A and B at 75$^{\prime \prime}$ and 45$^{\prime \prime}$ respectively to delineate their approximate infrared boundaries. The $\rm{3.6\mu m}$ observation shows the stellar content of G5.89-0.39 is highlighted in yellow.

We describe our Chandra observations and analysis in \S\ref{sec:data}. The X-ray source detection and spectral analysis is discussed in \S\ref{sec:xraysources}. \S\ref{sec:extended} discusses our search for unresolved point sources. \S\ref{sec:energetics} looks at the inferred energetics of G5.89-0.39 and possible connections to HESSJ1800-240B. The overall conclusions are given in \S\ref{sec:conc}.

\section{X-Ray Observations, Data Reduction, \& Source Detection}
\label{sec:data}
\subsection{Chandra Observation}
Our Chandra X-ray Observatory \citep{Weisskopf:2002aa} observation (Obs ID 10997; July 20-21, 2010 of $\rm{\sim78.8ks}$ duration) employed the ACIS instrument \citep{Garmire:2003aa} I-array CCDs in Timed Exposure Faint telemetry mode. The bore-sight of our observation was positioned so that the four CCDs arranged in a square (CCD IDs 0, 1, 2, \& 3) adequately covered the HII complex G5.89-0.39. The 2 CCDs in the S array (CCD IDs 6, \& 7) were also included. Over 200 X-ray sources across the FoV were identified in a preliminary analysis of our observation with some sources clearly clustering around G5.89-0.39 \citep{Rowell:2010aa}.

\subsection{Data Reduction}
The Chandra observation of G5.89-0.39 was reduced using the data reduction tools in CIAOv4.4 \citep{Fruscione:2006aa}. We have followed the recommended data reduction processes, 
starting with the pre-analysis step \texttt{repro}. \texttt{Repro} corrects for dithering in order to retain the sub pixel resolution of a Chandra observation and creates an \texttt{event=2} fits file for analysis. 

We employed the use of four energy bands for our study; a full energy band of 0.3-10\,keV, and three narrow bands of 0.3-2.5\,keV (soft), 2.5-5\,keV (medium), and 5-10\,keV (hard). The narrow bands were chosen to separate X-ray emission heavily absorbed by photoelectric absorption (the hard band) due to the expected high column density of HII regions, and provide an unbiased census of soft and hard sources in the FoV. A more restricted energy band of 0.1-2\,keV is also processed for comparisons with ROSAT flux and luminosities of X-ray sources. 

Each energy band image was processed individually to eliminate flares using the \texttt{deflare} command in CIAOv4.4. This process uses a clipping where any events outside of 3$\sigma$ are taken out of the observation. Following this step, the images represent: 78.2\,ks, 78.2\,ks, 78.8\,ks, and 78.2\,ks, for the full, hard, medium and soft bands respectively. We have also used \texttt{deflare} on the original event=2 image ahead of our spectral analysis discussed in \S\ref{sec:spect}. For each energy band image a PSF (point spread function) map and an exposure map were created for source detection and spectral analysis.

\subsection{X-ray Source Detection }
\label{sec:wav}
We employed \texttt{wavdetect}, a source detection algorithm, to identify possible X-ray sources in our Chandra observation. \texttt{wavdetect} is ideal for detecting closely-spaced point sources and small-scale extended diffuse emission. Preliminary analysis with source detection algorithms within CIAOv4.4 determined that \texttt{wavdetect} was able to identify a third more X-ray sources in the closely packed HII complex over the other often-used algorithm, \texttt{celldetect}.

\noindent Values are extracted in two steps by \texttt{wavdetect} for possible X-ray sources: 
\begin{enumerate}
\item The algorithm detects probable source pixels within the image and correlates them with a Mexican hat wavelet of different scale sizes, 1, 2, 4, 8, and 16 pixels. The detection significance value was set to $4.7\,\sigma$ above background fluctuations, which is equivalent to a $10^{-6}$ chance probability. 
\item A Table of information listing each source and the information extracted for each source, such as position, net photon counts, net photon count rate, significance, and PSF ratio (intrinsic radius of the source in PSF units) is generated. 
\end{enumerate}

Source detections were performed on all four of our energy bands resulting in 268 individual source detections made by \texttt{wavdetect}. Running the source detection on each energy band individually allowed for comparisons between them to identify the same source in multiple bands. We then compiled the source detections from each energy band to make a single master list of sources with the appropriate values from each energy band where a source is identified in multiple energy bands. We employed a spatial error of 1" to 2", depending on local PSF, for the position of a source when compiling the master list. 

Our analysis is based on the assumption that an X-ray point source is more likely to be stellar in origin if the spectral emission is across multiple energy bands. For this reason our first condition for a source to enter our final list is that it must be detected in at least two of our four energy bands. The second condition looks at the net photon counts and requires that a source must have at least 10 net photon counts in one of the energy bands it is detected in. This net photon count cut is on top of the 4.7$\sigma$ threshold placed on the source detection algorithm and is to remove the weakest of sources from our final list. The final condition we employ is to remove sources that are too close to the edge ($<5$ pixels) of the CCDs or near any discrepancies in the observation. A discrepancy is particularly prominent in CCD 2 where several rows of pixels are less sensitive than the rows around them.

\begin{figure*}
  \centering
  \includegraphics[width=0.875\textwidth]{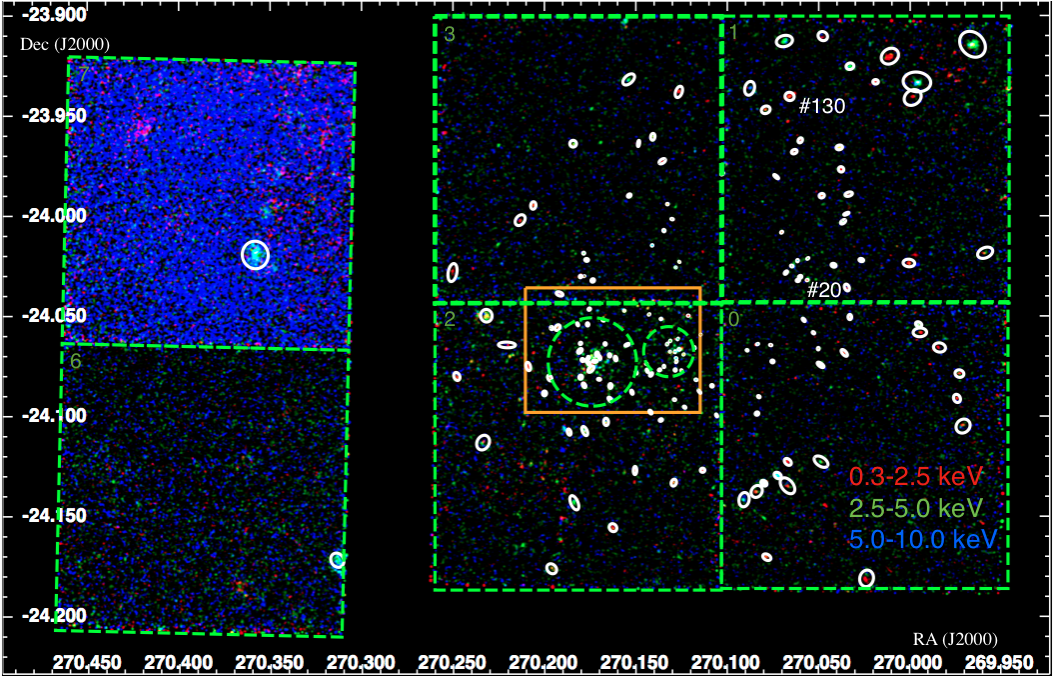}\\
  \includegraphics[width=0.875\textwidth]{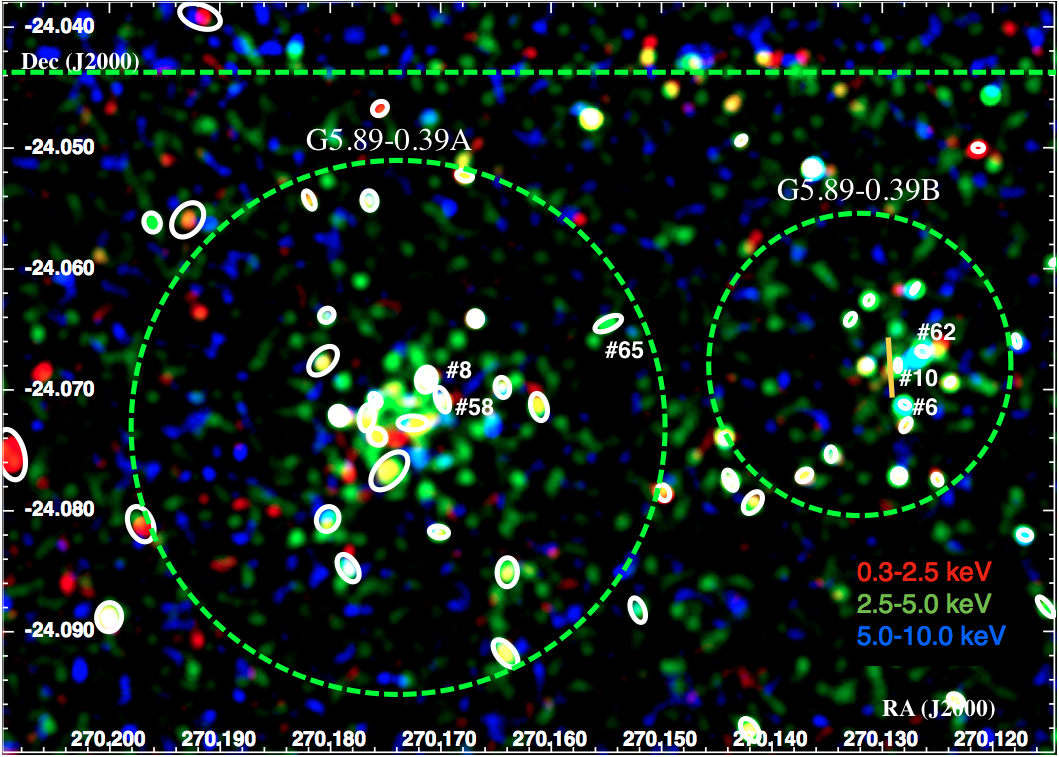}
  \caption{Chandra ACIS-I photon events (counts/pixel/cm$^{2}$/s) in our three energy bands, soft (0.3-2.5 keV) in red, medium (2.5-5 keV) in green and hard (5-10 keV) in blue. 
    White circles indicate some of the 159 identified X-ray sources. Dashed green circles indicate the two HII regions G5.89-0.39A and B. The top panel shows the entire FoV of our observation. 
    The bottom panel shows a zoom in on the HII complex G5.89-0.39. The numbered sources are the brightest of the sources detected in our observation. The yellow line indicates the possible molecular outflow from Feldt's star \citep{Feldt:2003vn}, which we believe is source \#10 (see \S \ref{sec:feldt}).}
  \label{fig:fullset}
\end{figure*}

A total of 159 X-ray sources are in our final list. 158 out of the 159 X-ray sources are detected in the full energy band. Source labelled \#159 was identified in the soft and medium energy band images only. Overall, $\sim 70\%$ of sources are identified in the soft energy band, $\sim 58\%$ of sources are identified in the medium energy band, and $\sim 15\%$ of sources are identified in the hard energy band. The complete catalogue of 159 sources is listed in \ref{ap:table} and Figure \ref{fig:fullset} shows a composite image of the three narrow energy band images with the 159 sources indicated by white circles/ellipses. A number of sources were found towards G5.89-0.39 and the changing sizes of the source regions highlight the varying PSF across the FoV. Our choice of the size of the radius of each HII region is discussed in \S\ref{sec:deabs1}.  

\begin{figure}
  \centering
  \includegraphics[width=0.5\textwidth]{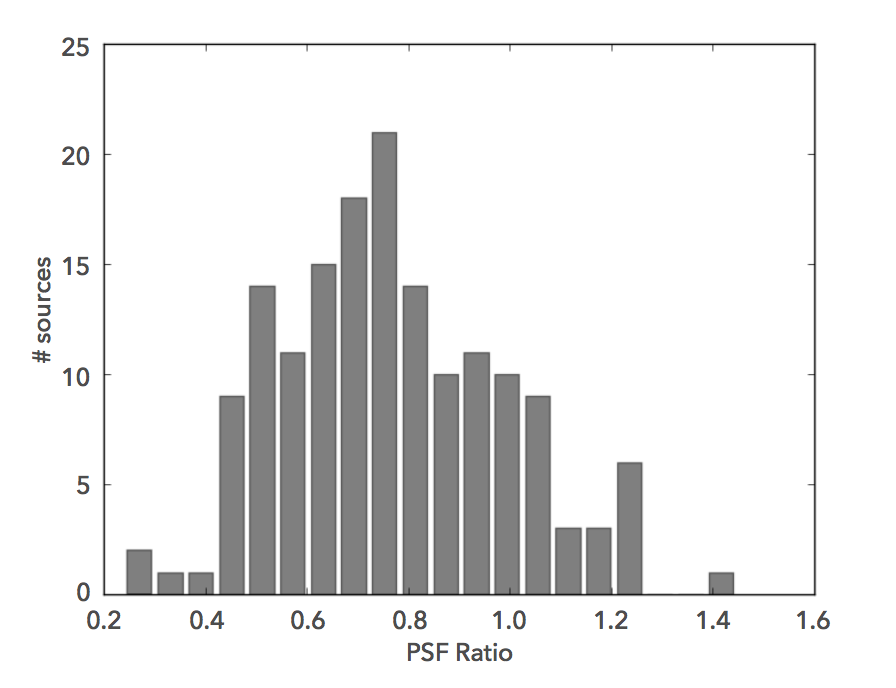}
  \caption{Distribution of the parameter PSFRATIO from the \texttt{wavdetect} algorithm for the 159 final X-ray sources.}
  \label{fig:psfratio}
\end{figure}

Our study of stellar sources also requires that our sources are point-like. To characterise the intrinsic spatial extent of the 159 sources, we considered the distribution of the parameter PSFRATIO produced by \texttt{wavdetect \footnote{See \url{http://cxc.harvard.edu/ciao/download/doc/detect\_manual/wav\_ref.html}}}. Figure \ref{fig:psfratio} shows the bulk of the PSFRATIO values are between 0.5 and 1.2 suggesting the sources can be as considered point-like. Source \#138, with the highest PSFRATIO of 1.44, is located in CCD\,0 outside the boundaries of G5.89-0.39A and B and has a single infrared counterpart (see Table.~\ref{tab:counter}), hence we consider it also point-like.

We have used the 2MASS point source catalogue and SIMBAD database to identify any possible counterparts in other wavelengths of our 159 X-ray sources. Using a search radius of $1*\rm{PSF}$ for each source we have identified 100 possible counterparts for 90 of our X-ray sources, Tables~\ref{tab:counter} and ~\ref{tab:counter2}. Source \#10 in our X-ray observation is a possible X-ray counterpart to Feldt's star \citep{Feldt:2003vn} which is discussed in more 
detail in \S\ref{sec:feldt}. Source \#62 overlaps the sub-millimeter continuum and line emission ring imaged by \citet{Hunter:2008fr} that is possibly blown out by Feldt's star.

The uncertainty of the matches is considered in two distinct groupings of sources; The 35 sources associated with G5.89-0.39, which this work concentrates on, and the entire source catalogue of 159 sources. Following the method of \citet{White:1991aa} we have calculated the by-chance coincidences with the 2MASS catalogue by shifting the right ascension (RA) of all identified X-ray sources in our observation by $\pm 10^{\prime\prime}$ and $\pm 20^{\prime\prime}$, to test a range of shifts, and searching each time in the 2MASS catalogue for matches to these shifted set of X-ray sources with the same spatial distribution of our X-ray sources. Table \ref{tab:coin} presents the results of searching for shifted-source matches using 1*PSF and 2*PSF search radii. We found an 11\% false-match rate for sources in G5.89-0.39 and 34\% false-match rate for the entire FoV with our 1*PSF search radius.

\begin{table*}
\centering
\begin{tabular}{ccccc}
\hline
\multicolumn{1}{c}{\# X-ray sources}&
\multicolumn{1}{c}{\# matched sources}&
\multicolumn{1}{c}{search radii}&
\multicolumn{1}{c}{-20$^{\prime\prime}$,-10$^{\prime\prime}$,10$^{\prime\prime}$,20$^{\prime\prime}$}&
\multicolumn{1}{c}{by-chance \%}\\
\hline
G5.89-0.39\\
\hline
35 & 15 & 1*PSF & 5,4,2,4, & 11\%\\
35 & 19 & 2*PSF & 14,13,10,19 & 40\%\\
\hline
entire FoV\\
\hline
159 & 97 & 1*PSF & 54,53,53,56 & 34\%\\
159 & 111 & 2*PSF & 99,102,95,105 & 63\%\\
\hline
\end{tabular}
\caption{Determination of the false-match rate with 2MASS counterparts by shifting the RA positions each X-ray source by $\pm$10" and $\pm$20".}
\label{tab:coin}
\end{table*}

\section{The Stellar Population of G5.89-0.39}
 \label{sec:xraysources}

\subsection{X-ray sources towards G5.89-0.39 }
\label{sec:deabs1}

Table \ref{tab:HIIs} lists the X-ray sources detected from the two distinct HII regions within G5.89-0.39. Each source is associated with a HII region if its position is within the boundaries of one of the regions based on the Spitzer $\rm{8\mu m}$ emission, see Table \ref{tab:radec}. We have calculated the observed and unabsorbed (see \S\ref{sec:lumdeabs}) luminosities, using a distance of 2\,kpc, for each source in the 0.3-10 keV energy band as well as listing the 0.1-2 keV luminosities for comparison with ROSAT derived luminosities. The observed X-ray luminosity, photon flux, and energy flux between 0.3 and 10 keV for each source was calculated via the CIAO task \texttt{eff2evt}. The energy flux was generated by summing over the event by event energy flux within each source and then converted to a luminosity assuming a distance of 2 kpc. This removed the need to assume a model spectrum and thus associated biases attached to such models. Table \ref{tab:HIIs} also lists the source number and RA/Dec position for each source in G5.89-0.39.

For each source we determined its likelihood of being variable in the X-ray flux. This was to determine if we have any flaring pre-main sequence (PMS) stars. To do this we used the CIAO command \texttt{glvary}\footnote{For more detailed explanation see \url{http://cxc.harvard.edu/csc/why/gregory_loredo.html}} using a Gregory-Loredi variability algorithm to determine if there is a time variability in the sources of G5.89-0.39. Of the 35 X-ray sources identified within G5.89-0.39 10 sources are considered to be variable. The three sources \#13, \#15, and \#49 have less than 15 net counts in our X-ray energy band across the whole observation so their variability is not certain. The sources \#43, \#44, \#57, and \#73 have between 25 and 30 net counts in our X-ray energy band over the entire observation. The three sources \#8, \#56, and \#68 have the highest net counts, $\sim80$, $\sim45$, $\sim39$ respectively. The light curves of these brightest sources show strong spikes of variable flux over the time of the observation. The seven sources with more than 15 net counts show a variable flux with time indicative of a PMS source. The remaining three are two few counts to investigate their variability. Of the three sources we were unable to test, only source \#15 has a counterpart in the 2MASS catalogue. Of the seven possible PMS stars two, sources \#8 and \#73, have 2MASS counterparts (see Table \ref{tab:counter} for counterpart IDs).

An analysis of all the X-ray sources identified in our Chandra observation found 17\% of all 159 sources were variable (variability index $>6$). For the HII complex G5.89-0.39 29\% of sources are variable. This shows that the sources in G5.89-0.39 are above the average in number of variable sources.

\begin{table}
\centering
\begin{tabular}{ccc}
\hline
\multicolumn{1}{c}{}&
\multicolumn{1}{c}{G5.89-0.39 A}&
\multicolumn{1}{c}{G5.89-0.39 B}\\
\hline
RA (J2000) & 270.174 & 270.132\\
Dec (J2000) & -24.073 & -24.068\\
Radius (arcsec) & 79.5 & 45\\
Radius (pc) & 0.8 & 0.4\\
\# Sources & 22 & 13\\
\hline
\end{tabular}
\caption{Positions and numbers of X-ray sources associated with the two HII regions and the boundary radius of each HII region (based on Spitzer $\rm{8\mu m}$ emission) 
  given as the radius in pc units assuming a distance of 2\,kpc.}
\label{tab:radec}
\end{table}

\begin{table*}
\setlength{\tabcolsep}{3pt}
\centering
\begin{tabular}{cccccc}
\hline
  \multicolumn{1}{c}{} &
  \multicolumn{1}{c}{}&
  \multicolumn{1}{c}{}&
  \multicolumn{2}{c}{Luminosities$^{\dagger}$} \\
  \multicolumn{1}{c}{Source \#} &
  \multicolumn{1}{c}{RA (J2000)}&
  \multicolumn{1}{c}{Dec (J2000)}&
  \multicolumn{1}{c}{$\rm{log \, L_{R}}$}&
  \multicolumn{1}{c}{$\rm{log \, L_{f}}$}&
   \multicolumn{1}{c}{X-ray Variability Index$^{\ddagger}$}
  \\
  &
  &
  &
  \multicolumn{1}{c}{($\rm{0.1-2.0\, kev}$)}&
  \multicolumn{1}{c}{($\rm{0.3-10\, keV}$)}
  \\
  
  &
  \multicolumn{1}{c}{(degrees)}&
  \multicolumn{1}{c}{(degrees)}&
  \multicolumn{1}{c}{($\rm{erg\,s^{-1}}$)}&
  \multicolumn{1}{c}{($\rm{erg\,s^{-1}}$)}
  \\
  \hline 
  4$\rm{^{B}}$ & 270.129 & -24.077 & 30.61 & 31.08 & 2\\ 
  6$\rm{^{B}}$ & 270.128 & -24.071 & - & 31.05 & 0\\ 
  7$\rm{^{B}}$ & 270.124 & -24.069 & 30.40 & 30.46 & 0 \\ 
  8$\rm{^{A}}$ & 270.171 & -24.069 & 31.50 & 31.36 & 8\\ 
  9$\rm{^{B}}$ & 270.131 & -24.068 & 30.69 & 30.77 & 2\\ 
  10$\rm{^{B}}$ & 270.129 & -24.068 & 29.76 & 31.32 & 1\\ 
  13$\rm{^{B}}$ & 270.127 & -24.062 & 30.11 & 30.55 & 7\\ 
  15$\rm{^{A}}$ & 270.168 & -24.052 & 30.66 & 30.46 & 6\\ 
  36$\rm{^{A}}$ & 270.164 & -24.092 & 31.07 & 31.31 & 1\\ 
  43$\rm{^{A}}$ & 270.164 & -24.085 & 30.78 & 31.39 & 6\\ 
  44$\rm{^{A}}$ & 270.178 & -24.085 & 30.97 & 31.76 & 8\\ 
  46$\rm{^{A}}$ & 270.17 & -24.082 & 30.84 & 31.44 & 0\\ 
  49$\rm{^{B}}$ & 270.125 & -24.077 & 30.28 & 30.23 & 6\\ 
  50$\rm{^{B}}$ & 270.137 & -24.077 & 30.68 & 30.53 & 2\\ 
  51$\rm{^{A}}$ & 270.175 & -24.077 & 31.36 & 31.97 & 0\\ 
  53$\rm{^{A}}$ & 270.176 & -24.074 & 31.14 & 31.07 & 0\\ 
  55$\rm{^{A}}$ & 270.177 & -24.072 & 30.87 & 31.54 & 0\\ 
  56$\rm{^{A}}$ & 270.179 & -24.072 & 30.80 & 31.41 & 7\\ 
  57$\rm{^{A}}$ & 270.161 & -24.071 & 30.84 & 31.55 & 6\\ 
  58$\rm{^{A}}$ & 270.17 & -24.071 & 31.77 & 31.47 & 0\\ 
  59$\rm{^{A}}$ & 270.176 & -24.071 & 30.72 & 31.20 & 0\\ 
  60$\rm{^{A}}$ & 270.164 & -24.07 & 30.55 & 31.41 & 2\\ 
  61$\rm{^{A}}$ & 270.181 & -24.068 & 30.85 & 31.66 & 0\\ 
  62$\rm{^{B}}$ & 270.126 & -24.067 & 30.45 & 31.05 & 2\\ 
  65$\rm{^{A}}$ & 270.155 & -24.065 & 31.78 & 31.06 & 1\\ 
  66$\rm{^{B}}$ & 270.133 & -24.064 & 30.47 & 30.26 & 1\\ 
  68$\rm{^{A}}$ & 270.167 & -24.064 & 31.00 & 31.64 & 6\\ 
  69$\rm{^{A}}$ & 270.18 & -24.064 & 30.77 & 31.41 & 1\\ 
  70$\rm{^{B}}$ & 270.131 & -24.063 & 29.85 & 30.51 & 0\\ 
  73$\rm{^{A}}$ & 270.177 & -24.054 & 30.73 & 31.09 & 6\\ 
  74$\rm{^{A}}$ & 270.182 & -24.054 & 30.54 & 30.28 & 0\\ 
  106$\rm{^{A}}$ & 270.18 & -24.081 & 30.98 & 31.31 & 0\\ 
  108$\rm{^{B}}$ & 270.135 & -24.075 & 30.47 & 31.33 & 1\\ 
  109$\rm{^{A}}$ & 270.173 & -24.073 & 30.79 & 31.61 & 1\\ 
  110$\rm{^{B}}$ & 270.128 & -24.073 & 30.30 & 30.56 & 1\\ 
  \hline
\end{tabular}
\caption{Un-absorbed luminosities for the X-ray sources associated with G5.89-0.39 assuming a distance of 2 kpc. A - Sources identified in G5.89-0.39A. 
  B - Sources identified in G5.89-0.39B. \\
  $\dagger$ - X-ray luminosities: R = ROSAT energy band (0.1-2 keV) f = full band (0.3-10.0 keV). \\
  Average error in the luminosity is $\sim 25 \%$ based on the net photon counts for each source as reported by \texttt{wavdetect}.\\
  $\ddagger$ - X-ray source variability. A variability index of 0-2 is not variable, index of 3-5 is possibly variable and a variability index $>$ 6 is indicative of a variable source.}
\label{tab:HIIs}
\end{table*}

\subsection{Spectral Analysis of X-ray Sources in G5.89-0.39}
\label{sec:spect}
Since most sources by themselves did not meet the minimum 10 counts per 10 energy bins for spectral analysis, we stacked all X-ray sources within G5.89-0.39 together to perform a co-stacked global spectral analysis.

Three different spectral models were fit to the stacked source spectra using the program XSPEC \citep[][]{Arnaud:1996fk}. The spectra were extracted from the original \texttt{event=2} image containing all photon events after using the CIAO command \texttt{deflare}. Since our observations were well within the influence of the diffuse Galactic Diffuse X-ray Emission (GDXE) \citep{Sawada:2012aa}, we did not employ the 
Chandra blank-sky (taken on extragalactic fields) files but instead chose local background regions in each of the ACIS-I CCDs for our spectral analysis.

An APEC thermal plasma model, non-thermal power-law model, and a thermal Bremsstrahlung model were all fit in turn with a photo-electric absorption model to the stacked spectra of sources. The photo-electric model is included to model the line-of-sight hydrogen column density, $\rm{N_{H} \, (cm^{-2})}$, towards the two HII regions. Table~\ref{tab:fits} presents the results of fitting the three models in terms of the fit parameters associated with each model and the model fit statistics, $\chi^2$/degrees of freedom.

\begin{table*}
\centering
\begin{tabular}{cccc}
\hline
\multicolumn{1}{c}{Model} &
\multicolumn{1}{c}{APEC} &
\multicolumn{1}{c}{Power Law} &
\multicolumn{1}{c}{Bremsstrahlung}\\
\hline
$\rm{N_{H}}$ [$\rm{10^{22}\,cm^{-2}}$] & $2.54 \pm \, 0.19$ & $2.83 \pm \, 0.26$ & $2.42 \pm \, 0.19$\\ 
kT [keV] & $4.52 \pm \, 0.82$ & & $5.32 \pm \, 1.02$\\
Abundance [solar units] & $0.34 \pm 0.17$ & & \\
Photon Index [$\Gamma$] & & $ 2.15 \pm \, 0.17$ & \\
Normalisation [$10^{-4}$] & $\rm{4.82 \pm 0.58^{\dagger}}$ & $2.08 \pm 0.51$$^{\ddagger}$& $1.51 \pm 0.19^{*}$ \\
$\rm{\chi^{2}/dof}$ & 50/47 & 57/48 & 54/48 \\
\hline\end{tabular}
\caption{Spectral models for X-ray sources of the HII Complex G5.89-0.39A and B fit in the 1-10 keV band. \\
$\dagger$ - normalisation for APEC model is equivalent to $\rm{10^{-14}/(4\pi (D_{A})^{2})\int n_{e}n_{H}dV\,cm^{-5}}$\\
${\ddagger}$ - power-law normalisation ($\rm{ph\,keV^{-1}\,cm^{-2}\,s^{-1}}$ at 1keV)\\
{*} - Bremsstrahlung normalisation is equivalent to $\rm{10^{-14}/(4\pi (D_{A})^{2})\,cm^{-5}}$}
\label{tab:fits}
\end{table*}

Adequate fits to the stacked spectra were achieved by all three models as show by the $\chi^2$/dof probabilities ($\sim10^{-1}$) listed in Table~\ref{tab:fits}. Figures~\ref{fig:mods1} and \ref{fig:mods2} presents the fits for each of the models with the combined photo-electric absorption model. The data are shown as crosses with error bars and model fits as the solid lines. The residuals of the model fits are presented below each graph in Figures~\ref{fig:mods1} and \ref{fig:mods2}.

\begin{figure*}
  \centering
      \includegraphics[width=0.75\textwidth]{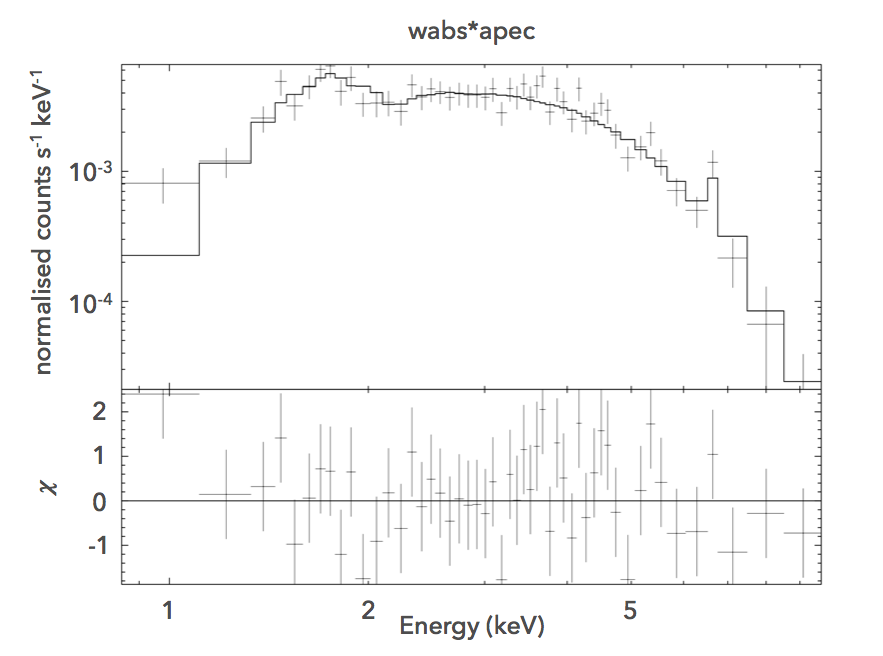} 
      \includegraphics[width=0.75\textwidth]{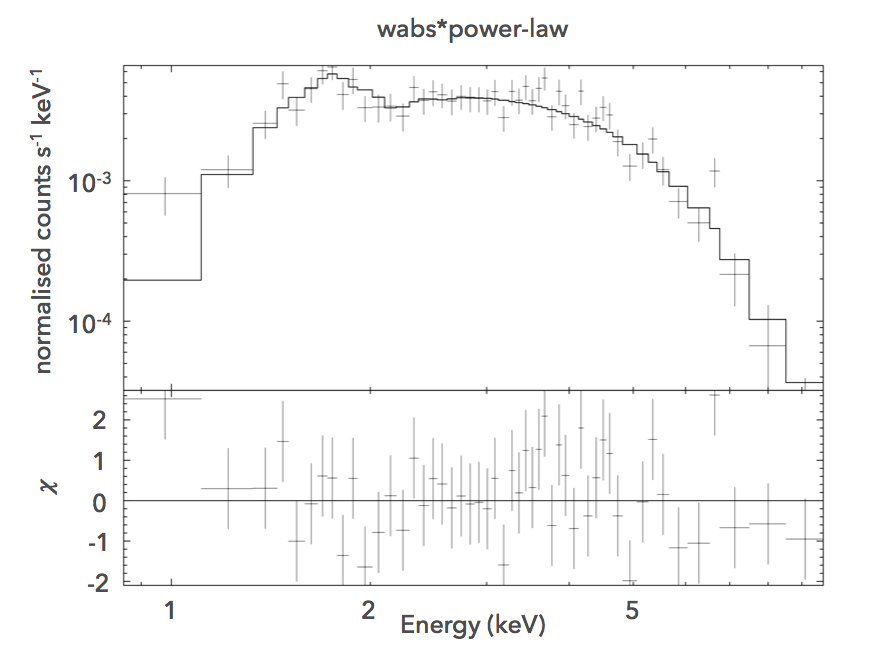} 
   
     \caption{(a) Spectral fits to the 35 X-ray sources identified in G5.89-0.39 with a thermal APEC model; (b) non-thermal power model. For all plots: Top panels show the spectra as crosses and 
        the model as the solid black line. Bottom panels show the residuals of each model fit. Continued with Fig.~\ref{fig:mods2}.}
         \label{fig:mods1}
\end{figure*}
   
  \begin{figure*}
  \centering
      \includegraphics[width=0.75\textwidth]{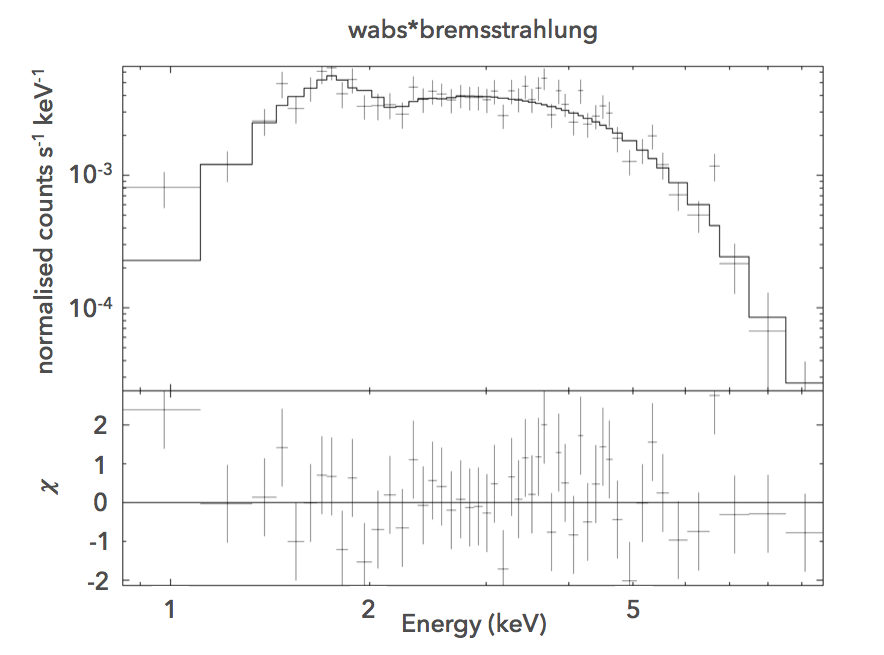} 
      \caption{Continuation of Fig.~\ref{fig:mods1} (c) thermal Bremsstrahlung model. Top panel shows the spectra as crosses and 
        the model as the solid black line. Bottom panel shows the residuals of each model fit.}
      \label{fig:mods2}
\end{figure*}

The model parameters listed in Table~\ref{tab:fits} can also be used to determine which model best represents our stacked spectra. Previous studies of G5.89-0.39 and other galactic SFRs provide some guidance concerning the parameters thermal temperature, abundance of metals, and hydrogen column density for our spectral analysis. The values from previous studies are summarised in Table~\ref{tab:vals}. 

We note that the line-of-sight column densities obtained for G5.89-0.39 in previous studies are inconsistent with each other due to the beam sizes of each survey of the area. However, studies on scales less than an arc-minute also obtained column densities over 10 times higher ($\rm{\sim10^{23}cm^{-2}}$) than those calculated by \citet{Nicholas:2012aa} \citep[][]{Hunter:2008fr}.

\begin{table*}
\centering
\begin{tabular}{ccccc}
\hline
\multicolumn{1}{c}{source/s} &
\multicolumn{1}{c}{kT [keV]} &
\multicolumn{1}{c}{abundance [$Z\odot$]} &
\multicolumn{1}{c}{$\rm{N_{H}}$ [$\rm{10^{22}\,cm^{-2}}$]} &
\multicolumn{1}{c}{reference}\\
\hline
G5.89-0.39A & & & $\rm{3.5}$ & \citet{Nicholas:2011aa}\\
G5.89-0.39B & & & $\rm{2.3}$& \citet{Nicholas:2011aa}\\
G5.89-0.39A and B & & & $\rm{\sim10}$ & \citet{Hunter:2008fr}\\
Class I protostar X-ray flares & $\sim$5 & & & \citet{Schulz:2001aa}\\
UCHII Sagittarius B2 & $\sim$6-10 & & & \citet{Takagi:2002zr}\\
UCHII W49A B-type sars & $\rm{\sim 7}$& 0.3 & & \citet{Tsujimoto:2006aa}\\
Massive star cluster 30 Doradus & 1\,-\,3 & $0.3$ & & \citet{Townsley:2006mz}\\
G5.89-0.39A and B & $4.52 \pm \, 0.82$ & $0.34 \pm 0.17$ & $2.54 \pm \, 0.19$ & this study\\
\hline\end{tabular}
\caption{Summary of values obtained by previous studies of G5.89-0.39 and other SFRs to compare to the values obtained through the fitting of the stacked X-ray spectra of sources in G5.89-0.39.}
\label{tab:vals}
\end{table*}

Stacking the spectra from both HII regions in G5.89-0.39 has complicated the comparison to previous studies. The parameters from our model fits to the stacked sources (Table \ref{tab:fits}) are consistent with the range of values obtained by previous studies (Table \ref{tab:vals}) encompassing a variety of HII regions and massive stellar clusters. 

We also found a marginally significant, $\sim2\sigma$, peak in the X-ray emission at 6.7\,keV. The 6.7\,keV peak is at the position of the Fe K$\alpha$ line and was revealed by the APEC model in our stacked spectra. This line is of particular interest in situations of high energy kinematics as it has been related to massive stars \citep[e.g.][]{anderson11}, HII regions \citep[e.g.][]{Takagi:2002zr}, and supernovae and SNRs \citep[e.g.][]{Helder:2012bh}. An earlier claim for Fe K$\alpha$ emission from W28 \citep{Rho:2002aa}, a probable nearby source of this line, has since been attributed to the GDXE (Galactic Diffuse X-Ray Emission) \citep{Sawada:2012aa,Zhou:2014aa}. Given the established presence of the K$\alpha$ iron in the GDXE 
which is in the raw background spectra of our observation, the 6.7 keV peak we detected is likely to come from the GDXE. Figure \ref{fig:kalph} shows the background emission from regions indicated in Figure \ref{fig:bkgs} as the red points. The background emission shows a peak around 6.7\,keV suggesting the emission to be from the GDXE and not a localised source within the HII complex.

\subsection{Un-absorbed Luminosities of G5.89-0.39 X-ray Sources}
\label{sec:lumdeabs}

We calculated the unabsorbed luminosity of each source in G5.89-0.39 assuming a thermal APEC model. Although the APEC, power-law, and bremsstrahlung models were all equally well fit, given the point-like nature of the X-ray sources, we assumed them to arise from thermal stellar processes and so used the APEC model in subsequent calculations. We made the conversions from photon events to energy flux using the CIAO command \texttt{eff2evt} and then used PIMMS with the optimised parameters of the APEC model fit to calculate the unabsorbed energy flux for each source. 

Using a distance of 2\,kpc we calculated the unabsorbed luminosities of each X-ray source in G5.89-0.39 in our full energy band and the ROSAT energy band. We include the luminosities converted to the ROSAT energy band of 0.1-2\,keV for comparison with \citet{Berghoefer:1997uq}. We also calculated the systematic errors in both energy bands using the extreme combinations of the fit APEC temperature and abundance values. The statistical errors propagated from the net counts of each source are twice that of the systematic errors, and are therefore the errors quoted in Table \ref{tab:HIIs}.

\begin{figure}
  \centering
  \includegraphics[width=0.5\textwidth]{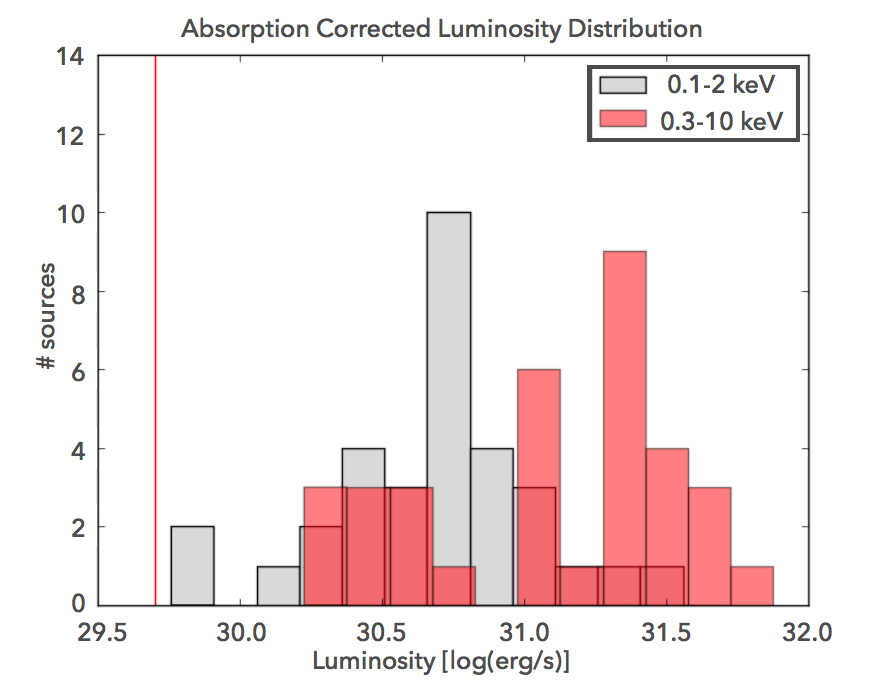}
  \caption{Unabsorbed luminosity distribution of X-ray sources in G5.89-0.39. Calculated over the energy band of 0.3-10 keV using \texttt{PIMMS v4.6a}, (red); Calculated over the energy band of 
    0.1-2 keV for comparisons with \citet{Berghoefer:1997uq} (grey). Red vertical line indicates the luminosity threshold at 4.7$\sigma$.}
  \label{fig:deabs1}
\end{figure}

Figure \ref{fig:deabs1} presents the 0.1-2\,keV (grey) and 0.3-10\,keV (red) luminosity distributions for G5.89-0.39. The mean unabsorbed luminosity (0.3-10\,keV) of the X-ray sources in G5.89-0.39A is $\rm{2.8\pm 0.6 \times 10^{31}\, erg\,s^{-1}}$ and G5.89-0.39B is $\rm{7.8\pm 1.8 \times 10^{30}\,erg\,s^{-1}}$. We note the {\tt wavelet} detection threshold of 4.7$\sigma$ corresponds to an unabsorbed luminosity (0.3-10\,keV) of $\rm{1.2\times 10^{30}\,erg\,s^{-1}}$, red line in Figure \ref{fig:deabs1}.

Due to the difference in the two HII regions comprising G5.89-0.39 we expect to see the observed difference in mean luminosity. G5.89-0.39A shows a higher luminosity distribution than G5.89-0.39B, most likely due to the age difference between the two regions. G5.89-0.39A is an older, more evolved SFR with a probable greater fraction of its molecular mass converted to stars. Whereas G5.89-0.39B is a younger SFR with probably more molecular gas compared to the number of stars. 

A comparison of our source luminosities (calculated for the ROSAT energy band) to a study by \citet{Berghoefer:1997uq} shows that our sources are consistent with B-Type stars, specifically B5-B0 spectral types, giving new insight into the stellar makeup of G5.89-0.39. Most sources appear to be consistently B-Type however, three sources have luminosities consistent with late O-type stars. These three sources, \#8, \#58 and \#65 all within G5.89-0.39A and labelled in Figure~\ref{fig:fullset} (bottom panel), are the most energetic sources we have identified in G5.89-0.39. We have found possible counterparts for \#8 and \#58 from the 2MASS point source catalogue but no counterpart for source \#65 was found. 


\subsection{Feldt's Star }
\label{sec:feldt}

We identified a possible counterpart to the suspected ionising source of G5.89-0.39B know as Feldt's star \citep{Feldt:2003vn}. Previous studies of Feldt's star have been able to determine its spectral type through bolometric luminosities as derived from the spectral energy distribution \citep{Faison:1998ul} or via the Lyman continuum photon budget \citep{Wood:1989ve, Kim:2003aa}. Each of the estimates of the spectral type are consistent, suggesting a spectral type O6. However, each of these estimates assumed that the source is in the centre of a spherical HII region and at a distance of $\rm{d=2.6\,kpc}$, greater than the 2\,kpc we are using for our Chandra study. The distance difference is enough to convert the O6 type star calculated by \citet{Faison:1998ul,Wood:1989ve,Kim:2003aa} to a B5-type star, consistent with \citet{Feldt:2003vn}.

\citet{Ball:1992pd} were the first to conclude that the ionising source must be off-centre based on their mid-infrared imaging, although the source itself had not yet been detected. \citet{Feldt:2003vn} was able to observe the ionising source using the Very Large Telescope (VLT) deriving a new distance of $\rm{d=1.9\,kpc}$, consistent with the distance we have adopted, which is based on studies of W28 and the surrounding molecular gas \citep{Aharonian:2008aa}. \citet{Hunter:2008fr} revealed a ring-like structure in sub-millimetre continuum and spectral bands confirming Feldt's star position was off-centre. 

Feldt's star is still the most likely source of ionisation in G5.89-0.39 despite the presence of molecular outflows possibly from an undetected protostar. One of the molecular outflows, extending out about 5\,arc-seconds in each direction, is indicated in Figure~\ref{fig:fullset} (bottom panel) by the yellow dashed line. The approximate orientation and positioning is from \citet{Sollins:2004aa,Puga:2006aa} as we don't detect the outflow in our observation. This outflow emanates from a region near Feldt's star. 

Our X-ray source \#10, which is coincident with the position of Feldt's star, has an estimated spectral type of B7-B5, based on its 0.1-2\,keV luminosity comparison with \citet{Berghoefer:1997uq}. The derived unabsorbed luminosity (0.1-2 keV) is $\rm{L=5.74 \pm 1.49 \times 10^{29} \, erg\, s^{-1}}$. 
 

\section{Unresolved Sources in G5.89-0.39}
\label{sec:extended}
After subtracting the stacked sources associated with both HII regions in G5.89-0.39, we found that some residual counts remained within the HII boundaries using nearby regions for local background estimates (see Table \ref{tab:netcounts} and Figure \ref{fig:bkgs} indicating the background regions). 

Unresolved point sources is a possible scenario for this residual X-ray emission. A spectral calculation of the luminosity of the residual X-ray emission was not possible due to the high background to residual counts ratio (see Table \ref{tab:netcounts}). Nevertheless a simplistic estimate of the residual X-ray luminosity was made by scaling the stacked source spectra luminosities by the ratios of the background net counts to the stacked source counts for the A and B components respectively (for A and B the ratios are 1.30 and 1.23). With this method, the residual luminosities convert to about 29 and 11 B-type stellar objects for G5.89-0.39A and B respectively.

\begin{table}
\centering
\begin{tabular}{ccc} \hline
           Regions        & G5.89-0.39A & G5.89-0.39B\\ \hline
Stacked Sources    & 656$\pm$29  & 247$\pm$17\\  
background         & 199         & 24        \\
Residual Emission  & 855$\pm$69  & 304$\pm$38\\  
background         & 3112        & 928       \\
\hline
\end{tabular}
\caption{Comparison of the 0.3-10\,keV net counts for the stacked resolved sources and residual emission within G5.89-0.39 A and B. The background counts used in the net counts calculation are also included.}
\label{tab:netcounts}
\end{table}

The residual unabsorbed luminosity of G5.89-0.39A, at 2\,kpc, is $\rm{3.6\times 10^{31}\, erg/s}$ and $\rm{9.6\times 10^{30}\, erg/s}$ for G5.89-0.39B. The unabsorbed luminosities have assumed the same APEC model as used for the individual detected source luminosities. For G5.89-0.39A we extracted the energy flux with \texttt{eff2evt} and then converted to an unabsorbed energy flux (for a distance of 2\,kpc) to obtain the residual unabsorbed X-ray emission. 

There is a possible total of 75 B-Type stellar sources in G5.89-0.39, taking into account the residual emission. Higher column densities towards the centres of each HII component in G5.89-0.39 may have caused our number to be underestimates. Overall, our estimates are broadly consistent with the estimate by \citet{Kim:2003aa} of the total stellar mass of high mass stars at about 700\,M$_\odot$ within G5.89-0.39 based on Lyman continuum brightness measurements.

G5.89-0.39 is towards the lower mass end of HII regions previously studied in the X-ray. The previous studies have focused on much larger systems with established clusters of B and O stars which generally have multiple HII components \citep[e.g.][]{Blum:1999gf,Moffat:2002aa,Knodlseder:2003ly,Tsujimoto:2006aa,Broos:2007,Kuhn:2013}. The comprehensive catalogue of \citet{Townsley:2014aa} for example summarised the X-ray sources from 11 massive SFRs and 30\,Doradus. Most of these SFRs have $>$1000 X-ray sources each of luminosity $> 10^{30}$\,erg/s. In terms of X-ray source numbers so far revealed, the grouping of UC HII regions in W49A \citep{Tsujimoto:2006aa} and the embedded young stellar cluster GGD\,27  \citep{Pravdo:2009aa} appear not too dissimilar to G5.89-0.39.

\section{Discussion of the Energetics of G5.89-0.39 and its Potential Connection to HESSJ1800-240B}
\label{sec:energetics}
Based on the un-absorbed luminosities inferred by our X-ray observations, we can estimate the combined kinetic power in G5.89-0.39, and 
hence evaluate its potential link to the TeV gamma-ray source HESSJ1800-240B. These power sources would need to meet 
that of the gamma-ray source which in the energy range 2 GeV to 3 TeV amounts to about $L_\gamma \sim 10^{34}$\,erg/s \citep{Aharonian:2008aa,Hanabata:2014aa}. 

An early B-type star with typical mass loss rate $\dot{M}\sim 10^{-7} M_\odot$~yr$^{-1}$ and terminal wind velocity $v_\infty \sim 500$km~s$^{-1}$ \citep{Cesarsky:1983,Lozinskaya:1992} 
would provide a wind luminosity ($L_w = 1/2 \dot{M} v^2_\infty$)
$\sim 10^{34}$\,erg/s. If we take the resolved 35 B-type stars in G5.89-0.39, and add the residual X-ray emission as stars, G5.89-0.39 could provide a total wind energy reaching $\sim 10^{36}$\,erg/s. 
In single stars, \citet{Weaver:1977} has suggested that up to 20\% of the wind energy could be transferred to 
kinetic energy. The additional scenarios discussed in \S\ref{sec:intro} involving stellar wind interactions in binary systems and/or in stellar 
clusters are also potentially involved. The protostellar outflow at the centre of G5.89-0.39B in particular, 
is well-known to have high energetics for such an object \citep{Sollins:2004aa,Watson:2007} at over $3.5\times 10^{46}$\,erg or $>10^{35}$\,erg/s over its dynamical lifetime of up to $\sim10^4$\,yr. We should also note the 
maser tracers (e.g. CH$_3$OH) detected towards G5.89-0.39A and our X-ray source \#20 about 0.1$^\circ$ north-west of G5.89-0.39B \citep{Nicholas:2012aa} (see Fig.~\ref{fig:ch3oh-cs10}) 
signal the ongoing presence of high mass star formation in several regions in our Chandra field of view and hence their potential to harbour protostellar outflows and winds.

All of these sources of kinetic energy could potentially set up multiple shocks which could then inject and accelerate cosmic-rays and electrons. Assuming the canonical 10\% of kinetic 
energy is transferred into accelerated particles (as per the paradigm for diffusive shock acceleration applied to SNRs), there is perhaps sufficent kinetic energy to power at least 
parts of the gamma-ray source HESSJ1800-240B. Concerning the type of particles responsible for the gamma-ray emission, the clear spatial overlap of the molecular gas and gamma-rays from HESSJ1800-240B 
strongly favours multi-TeV cosmic-rays as the preferred parent particles. Here, the cosmic-rays collide with the gas, creating gamma-rays from the decay of $\pi^\circ$ particles 
produced in these collisions. 

The alternative (or additional) avenue is via multi-TeV electrons,
which would involve X-ray synchrotron emission on the magnetic fields in the region, plus inverse-Compton emission in the TeV gamma-ray
band (up-scattering soft photon fields such as the cosmic microwave background and local IR emission). The great difficulty in fitting the broad GeV-TeV gamma-ray spectra of HESSJ1800-240B and the other gamma-ray sources surrounding the 
W28 SNR with electron-dominated models \citep[e.g.][]{Fujita:2009aa} is seen as further evidence for a cosmic-ray dominance. The lack of any evidence for 
non-thermal X-ray emission, probing for TeV electrons, in our Chandra observations would seem to support this. Given the likely dominance of thermal hot gas as the source of any extended X-ray emission in between the massive stars, 
this does not completely rule out the possibility of non-thermal emission.

Whether the accelerated particles actually reach the necessary multi-TeV energies is another question. \citet{Gusdorf:2015} have recently looked at the potential for particle acceleration in the inner regions 
($<$0.05\,pc) of G5.89-0.39B. They concluded that maximum particle energies may be of order 5\,GeV, too low to explain any TeV gamma-ray
emission if acceleration is directly governed by protostellar outflows. However, \citet{Gusdorf:2015} and several other studies \citep[e.g.][]{Domingo:2006,Bosch-Ramon:2010aa,Araudo:2014aa,Padovani:2016aa} 
have suggested scenarios involving the fast stellar winds from central stars driving the outflows or that of several other nearby massive stars are likely to provide the ideal conditions for cosmic-rays to reach 
TeV energies. Given the young age of G5.89-0.39 ($<10^6$\,yr), we may not realistically expect any additional particle acceleration due to so far undetected SNRs and/or pulsars within this HII complex.

Moreover, within HII regions, electron energies are expected to be limited to GeV energies by the strong synchrotron losses in the intense magnetic fields (of order 1\,mG as determined by \citet{Tang:2009} for
G5.89-0.39B), and inverse-Compton 
losses on the intense IR and optical/UV photon fields. The latter in fact may dominate over synchrotron losses \citep[e.g.][]{Reimer:2007}. The same radiative losses for electrons would also limit the size of any synchrotron X-ray and gamma-ray 
inverse-Compton emission to the inner sub-parsec regions of the HII regions. Unfortunately, probing such GeV electrons via non-thermal radio emission is often difficult due to the generally overwhelming dominance 
of thermal radio emission, and via GeV gamma-rays since the angular resolution of such measurements (FWHM $\sim$30--60 arc-minutes between 1-10\,GeV for Fermi-LAT) is well beyond the size scale of many HII regions.

In view of our results that now provide firm estimates of 
the stellar content and the X-ray energy budget in both HII region components of G5.89-0.39, these particle acceleration and gamma-ray production models can now be revisited with more precise inputs. At present the best gamma-ray angular 
resolution (FWHM $\sim$10 arc-min via H.E.S.S) is insufficient to distinguish potential gamma-ray components within HESSJ1800-240B. Higher resolution gamma-ray imaging on arc-minute scales towards HII regions such as this, with the 
forthcoming Cherenkov Telescope Array \citep{Bednarek:2013}, will greatly help in determining if there are any locally produced gamma-ray components in this region that may be generated by G5.89-0.39. Importantly, arc-minute gamma-ray 
resolution in particular will be able to exploit the energy-dependent diffusion of cosmic-rays and electrons from acceleration regions through dense molecular gas, which could impart rather specific gamma-ray spectral 
profiles across it (see e.g. \citet{Gabici:2007aa,Maxted:2012} for a discussion).

\section{Conclusions}
\label{sec:conc}

We have used the Chandra X-ray observatory to reveal the stellar content of the HII complex G5.89-0.39 found towards the gamma-ray source HESSJ1800-240B and dense molecular gas. The conclusions from
our work are as follows.

Using the ACIS-I CCD array, our observation ($\sim$78\,ks) reveals 159 resolved X-ray sources across the field of view. 
35 of these X-ray sources are found within the infrared-defined boundaries (0.8\,pc and 0.4\,pc radii) of the two components of G5.89-0.39A and B. Of these 35 sources, 22 are 
associated with the HII region G5.89-0.39A and 13 are associated with ultra-compact HII region G5.89-0.39B. The unabsorbed X-ray luminosities (0.1-2\,keV) of the sources associated 
with the HII regions are of order $10^{30.5}$\,erg/s suggesting they 
are early B-type stars, with several higher luminosity sources being possible O-type stars. Their stacked energy spectra is well-fit with a single thermal plasma APEC model with kT$\sim$5\,keV, 
metal abundance 0.35\,Z$_\odot$ and column density N$_{\rm H}=2.6\times10^{22}$\,cm$^{-2}$ (A$_{\rm V}\sim 10$). 

We have also identified a possible X-ray counterpart to the suggested ionising source in G5.89-0.39B known as Feldt's star. Its unabsorbed luminosity (0.1-2\,keV) of $\sim 6\times 10^{29}$\,erg/s is suggestive of a B7-B5 star, consistent 
with the lower limit on spectral type determined by \citet{Feldt:2003vn}. Allowing for an underestimate in the column density towards the centres of the HII regions 
(due to it being determined over a wide area using stacked source spectra), the luminosities and spectral types for some of the resolved sources are also possibly underestimated.

The residual (source-subtracted) X-ray emission encompassing G5.89-0.39A and B out to their boundaries was also examined and found to be about 30\% and 25\% larger than their respective stacked source luminosities.  
Assuming this residual emission is from unresolved stellar sources, the total B-type-equivalent stellar content in G5.89-0.39A and B would be about 75 stars. These estimates are consistent with 
the total mass of high mass stars of about 700\,M$_\odot$ determined by \citet{Kim:2003aa}.

The potential stellar population could provide wind energies reaching $\sim 10^{36}$\,erg/s. Factoring in efficiencies for converting wind energy into accelerated particles, this could account for
a part of the GeV to TeV gamma-ray emission in the source HESSJ1800-240B. Future arc-minute resolution gamma-ray imaging of this region with the Cherenkov Telescope Array \citep{Vercellone:2014aa} could help to reveal and disentangle 
potential gamma-ray components powered by the G5.89-0.39 HII complex.

The HII Complex G5.89-0.39 is at the lower end of total embedded stellar mass for the population of HII regions studied in X-rays. Previous studies such as \citet{Blum:1999gf,Moffat:2002aa,Knodlseder:2003ly,Tsujimoto:2006aa,Broos:2007,Kuhn:2013,Townsley:2014aa} found the number of O-Type stars to be much greater than the number we have identified for our region. \citet{Townsley:2014aa} found some SFRs with $>1000$ X-ray sources. However G5.89-0.39 does appear similar to the UC HII regions in W49A \citep{Tsujimoto:2006aa} and young stellar cluster in GGD\,27 \citep{Pravdo:2009aa}. Our study of the stellar census of G5.89-0.39 paves the way for a more detailed consideration of the potential for multi-TeV particle acceleration.

\section*{Acknowledgments}
This research has made use of the SIMBAD database, operated at CDS, Strasbourg, France. We would also like to thank Dr Jasmina Lazendic-Galloway for her insights into X-ray spectral analysis using XSPEC.

\bibliographystyle{mn2e.bst}
\bibliography{bibliography}
\clearpage
\appendix

\section{Comparisons with Mopra CH$_3$OH-I Maser and CS(1-0) Line Emission}
\begin{figure*}
  \centering
  \includegraphics[width=0.7\textwidth]{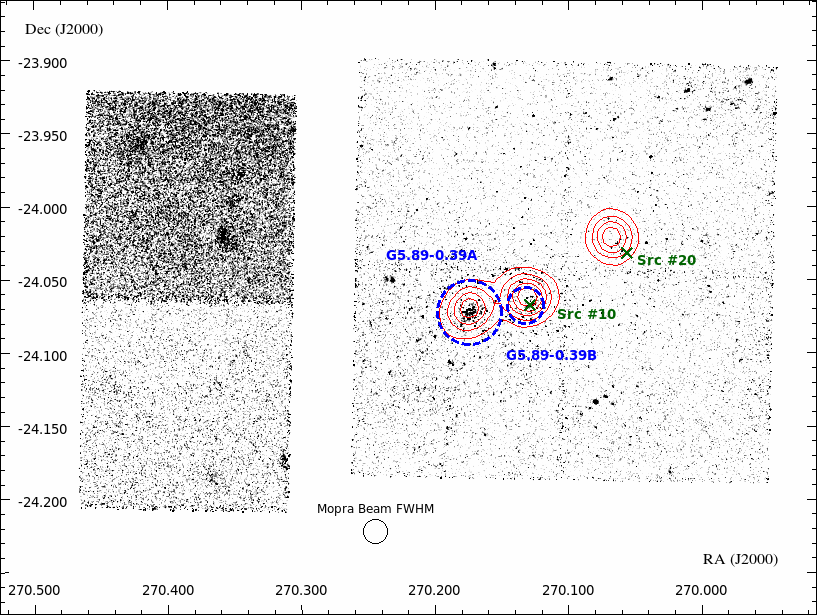}
  \includegraphics[width=0.7\textwidth]{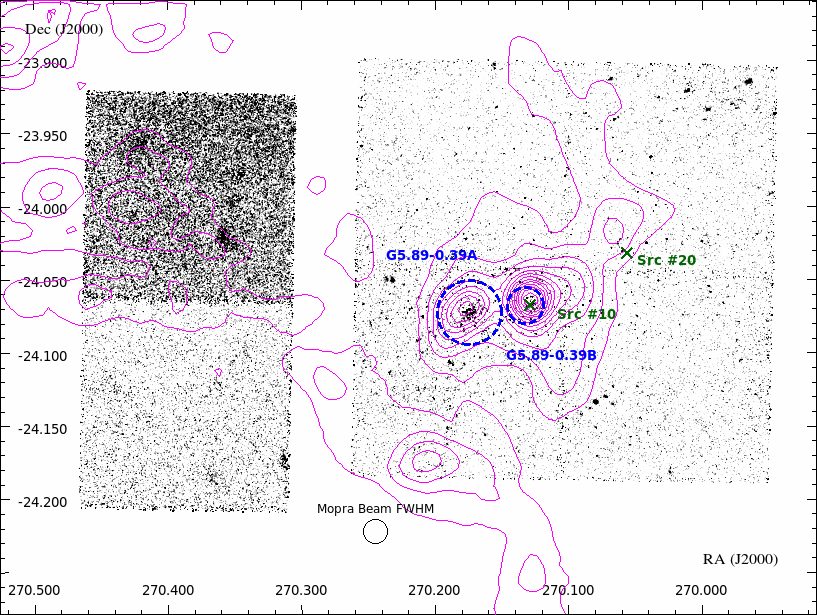}
  \caption{ACIS-I images (counts/pixel/cm$^2$/sec) with contours of: (Top) CH$_3$OH-I 44.1\,GHz from 1 to 6 K\,km/s in 6 levels (red) integrated from 0 to 15\,km/s; 
    (Bottom) CS(1-0) 49.0\,GHz from 2 to 28 K\,km/s in 14 levels (magenta) integrated from -5 to 25\,km/s. In both panels the locations of G5,89-0.39A, B, X-ray source \#10 (Feldt's star)
    and source \#20 are shown.}
  \label{fig:ch3oh-cs10}
\end{figure*}

Here, Fig.~\ref{fig:ch3oh-cs10} compares the ACIS-I image with Mopra observations \citep{Nicholas:2012aa} of the  CH$_3$OH-I Maser 44.1\,GHz and CS(1-0)  49.0\,GHz spectral line emission.

\section{Counterparts to X-ray Sources}
 \label{ap:counter}
A total of 90 of the identified X-ray sources have possible counterparts in the SIMBAD database or the 2MASS point source catalogue. 
Table \ref{tab:counter} and \ref{tab:counter2} summarises the possible counterparts to these X-ray sources. 

\begin{table*}
\setlength{\tabcolsep}{5pt}
\centering
\begin{tabular}{cccc}
\hline
  \multicolumn{1}{c}{Source \#} &
  \multicolumn{1}{c}{Possible Counterpart} &
  \multicolumn{1}{c}{Distance$^{\dagger}$}&
   \multicolumn{1}{c}{Source}\\
   \multicolumn{1}{c}{} &
  \multicolumn{1}{c}{Identifier} &
  \multicolumn{1}{c}{(arcsec)}&
   \multicolumn{1}{c}{type}\\
\hline 

$1$ & 2MASS J18001954-2408002$^{1}$ & 4.76 & IR source\\

$5^{B}$ & 2MASS J18003463-2404257$^{1}$ & 0.69 & IR source\\

$6^{B}$ & [BE83] IR 005.88-00.40$^{2}$ & 2.67 & IR source \\ 

7$^{B}$ & 2MASS J18002972-2404098$^{1}$ & 0.35 & IR source\\

8$^{B}$ & 2MASS J18004112-2404087$^{1}$ & 0.34 & IR source\\

$10^{B}$&2MASS J18003084-2404046$^{2}$&0.44 & IR source\\

15$^{A}$ & 2MASS J18004027-2403081$^{1}$ &0.36 & IR source\\

16 & 2MASS J18003272-2403063$^{1}$ & 0.42 & IR source\\

17 & 2MASS J18002914-2403003$^{1}$ & 0.35 & IR source\\


19 & 2MASS J18003754-2402514$^{1}$ & 0.39 & IR source\\

23 & 2MASS J18001000-2401292$^{1}$ & 0.61 & IR source\\

24 & 2MASS J18003031-2401232$^{1}$ & 0.10 & IR source\\

27 & 2MASS J18001685-2400302$^{1}$ & 0.32 & IR source\\

30 & 2MASS J18003184-2359396$^{1}$ & 0.49 & IR source\\

32 & 2MASS J18001725-2407498$^{1}$ & 5.29 & IR source\\



35 & 2MASS J18002959-2405444$^{1}$ & 0.33 & IR source\\

36$^{A}$ & 2MASS J18003932-2405310$^{1}$ & 1.33 & IR source\\

37 & 2MASS J18003094-2405310$^{1}$ & 0.16 & IR source\\

38 & 2MASS J18001986-2405254$^{1}$ & 0.22 & IR source\\

39 & 2MASS J18004797-2405201$^{1}$ & 0.97 & IR source\\

46$^{A}$ & 2MASS J18004084-2404544$^{1}$ & 0.21 & IR source\\

48 & 2MASS J18003595-2404423$^{1}$ & 0.48 & IR source\\

51$^{A}$ & 2MASS J18004191-2404359$^{1}$ & 0.43 & IR source\\

52 & 2MASS J18001169-2404288$^{1}$& 0.90 & IR source\\

54 & 2MASS J18001814-2404220$^{1}$ & 1.02 & IR source\\

57$^{A}$ & 2MASS J18003870-2404169$^{1}$ & 0.33 & IR source\\

58$^{A}$ & 2MASS J18004080-2404161$^{1}$ & 1.17 & IR source\\

$62^{B}$&2MASS J18003033-2404005$^{1}$&0.19 & IR source\\
&[HBI2008] SMA1$^{4}$&0.60 & sub mm source\\
& [BE83] Maser 005.89-00.39$^{2}$ & 0.94 & OH/IR star\\


70$^{B}$ & 2MASS J18003148-2403458$^{1}$ & 0.56 & IR source\\

71 & 2MASS J18001263-2403307$^{1}$ & 0.77 & IR source\\

72 & 2MASS J18004706-2403224$^{1}$ & 0.51 & IR source\\

73$^{A}$ & 2MASS J18004242-2403165$^{1}$ & 1.14 & IR source\\

74$^{A}$  & 2MASS J18004383-2403149$^{1}$ & 2.20 & IR source\\

75 & 2MASS J18000789-2403024$^{1}$ & 0.21 & IR source\\

\hline\end{tabular}
\caption{Possible counterparts from SIMBAD or 2MASS point source catalogue within $2*\rm{PSF}$ of numbered X-ray sources. Correlation with non point like objects 
  have been excluded. $^{\dagger}$ - Distance between centroid of X-ray source identified by \texttt{wavdetect} and central position of possible counterpart as 
  defined in the SIMBAD database or 2MASS point source catalogue. $^{A}$- identified within G5.89-0.39A, $^{B}$- identified within G5.89-0.39B.
$^{1}$ - \citet{Cutri:2003kx} 2MASS All Sky Catalog of point sources.
$^{2}$ - \citet{Braz:1983fj} Catalogue of non stellar molecular maser sources and their probable infrared counterparts.
$^{3}$ - \citet{Benjamin:2003yq} GLIMPSE. I. An SIRTF legacy project to map the inner galaxy.
$^{4}$ - \citet{Hunter:2008fr} Subarcsecond submillimeter imaging of the ultracompact H II region G5.89-0.39.}
\label{tab:counter}
\end{table*}

\begin{table*}
\setlength{\tabcolsep}{5pt}
\centering
\begin{tabular}{cccc}
\hline
  \multicolumn{1}{c}{Source \#} &
  \multicolumn{1}{c}{Possible Counterpart} &
  \multicolumn{1}{c}{Distance$^{\dagger}$}&
   \multicolumn{1}{c}{Source}\\
   \multicolumn{1}{c}{} &
  \multicolumn{1}{c}{Identifier} &
  \multicolumn{1}{c}{(arcsec)}&
   \multicolumn{1}{c}{type}\\
\hline

76 & 2MASS J18004212-2402483$^{1}$ & 0.23 & IR source\\

84 & 2MASS J18000643-2401194$^{1}$ & 0.58 & IR source\\

86 & 2MASS J18000833-2359566$^{1}$& 0.49 & IR source\\

88 & 2MASS J18001762-2358491$^{1}$ & 1.00 & IR source\\

89 & 2MASS J18002465-2358449$^{1}$ & 0.37 & IR source\\

90 & 2MASS J18003257-2358223$^{1}$ & 0.36 & IR source\\

91 & 2MASS J18000933-2357578$^{1}$ & 0.52 & IR source\\

92 & 2MASS J18003562-2357504$^{1}$ & 0.60& IR source\\

93 & 2MASS J18001437-2357440$^{1}$ & 1.12 & IR source\\

94 & 2MASS J18003032-2356160$^{1}$ & 1.87 & IR source\\

95 & 2MASS J17595181-2354517$^{1}$ & 0.31 & IR source\\

96 & 2MASS J18003898-2409222$^{1}$ & 1.48 & IR source\\

97 & 2MASS J18002019-2408165$^{1}$ & 0.85 & IR source\\

98 & 2MASS J18003099-2407598$^{1}$ & 0.81 & IR source\\

100 & 2MASS J18001606-2407227$^{1}$ & 0.12 & IR source\\

101 & 2MASS J18004477-2406302$^{1}$& 1.06 & IR source\\

102 & 2MASS J18004267-2406300$^{1}$ &2.89 & IR source\\

103 & 2MASS J18003991-2406122$^{1}$ & 1.45 & IR source\\

105 & 2MASS J18004731-2404527$^{1}$ & 0.86 & IR source\\

107 & 2MASS J18005012-2404325$^{1}$ & 1.44 & IR source\\

109$^{A}$ & 2MASS J18004132-2404216$^{1}$ & 1.51 & IR source\\

110$^{B}$ & 2MASS J18003071-2404225$^{1}$ & 0.09 & IR source\\

111 & 2MASS J18000865-2404068$^{1}$& 0.21 & IR source\\

112 & 2MASS J18005269-2403533$^{1}$& 3.88 & IR source\\

113 & 2MASS J18004630-2403205$^{1}$ & 0.86 & IR source\\

114 & 2MASS J17595858-2403201$^{1}$ & 6.18 & IR source\\


116 & 2MASS J18001396-2403054$^{1}$ & 1.71 & IR source\\

118 & 2MASS J18010006-2401396$^{1}$ & 2.76 & IR source\\

119 & 2MASS J18000012-2401249$^{1}$ & 0.56 & IR source\\

121 & 2MASS J18000857-2400096$^{1}$ & 3.49 & IR source\\

122 & 2MASS J18005120-2400089$^{1}$ & 0.76 & IR source\\

123 & 2MASS J18004946-2359407$^{1}$ & 1.19 & IR source\\

125 & 2MASS J18000812-2359215$^{1}$ & 0.80 & IR source\\

126 & 2MASS J18000907-2358371$^{1}$ & 0.24 & IR source\\

128 & 2MASS J18003375-2357386$^{1}$ & 0.64 & IR source\\

129 & 2MASS J18001900-2356498$^{1}$ & 0.45 & IR source\\

130 & 2MASS J18001620-2356215$^{1}$ & 6.72 & IR source\\ 
& SSTGLMC G005.9677-00.2810$^{3}$&1.05 & young stellar object candidate\\

\hline\end{tabular}
\caption{(continuation of Table B7) Possible counterparts from SIMBAD or 2MASS point source catalogue within $2*\rm{PSF}$ of numbered X-ray sources. 
  Correlation with non point like objects have been excluded. $^{\dagger}$ - Distance between centroid of X-ray source identified by \texttt{wavdetect} 
  and central position of possible counterpart as defined in the SIMBAD database or 2MASS point source catalogue. $^{A}$- identified within G5.89-0.39A, $^{B}$- identified within G5.89-0.39B.
  $^{1}$ - \citet{Braz:1983fj} Catalogue of non stellar molecular maser sources and their probable infrared counterparts.
  $^{2}$ - \citet{Cutri:2003kx} 2MASS All Sky Catalog of point sources.
  $^{3}$ - \citet{Benjamin:2003yq} GLIMPSE. I. An SIRTF legacy project to map the inner galaxy.
  $^{4}$ - \citet{Hunter:2008fr} Subarcsecond submillimeter imaging of the ultracompact H II region G5.89-0.39.}
\label{tab:counter2}
\end{table*}

\begin{table*}
\setlength{\tabcolsep}{5pt}
\centering
\begin{tabular}{cccc}
\hline
  \multicolumn{1}{c}{Source \#} &
  \multicolumn{1}{c}{Possible Counterpart} &
  \multicolumn{1}{c}{Distance$^{\dagger}$}&
   \multicolumn{1}{c}{Source}\\
   \multicolumn{1}{c}{} &
  \multicolumn{1}{c}{Identifier} &
  \multicolumn{1}{c}{(arcsec)}&
   \multicolumn{1}{c}{type}\\
\hline

131 & 2MASS J17595931-2355590$^{1}$ & 2.66 & IR source\\

132 & 2MASS J18003710-2355563$^{1}$& 2.28 & IR source \\

133 & 2MASS J18000275-2355146$^{1}$ & 1.56 & IR source \\

134 & 2MASS J18001661-2354434$^{1}$ & 3.18 & IR source \\

135 & 2MASS J18000580-2410532$^{1}$ & 1.90 & IR source \\

136 & 2MASS J18001891-2410142$^{1}$ & 1.34 & IR source \\

137 & 2MASS J18004399-2408380$^{1}$ & 2.55 & IR source \\

138 & 2MASS J18001569-2408057$^{1}$ & 5.47& IR source \\

139 & 2MASS J18002719-2407381$^{1}$ & 0.93 & IR source \\

141 & 2MASS J18005601-2406472$^{1}$& 1.53 & IR source \\

142 & 2MASS J17595303-2406160$^{1}$ & 2.18 & IR source \\

143 & 2MASS J17595369-2405282$^{1}$ & 1.89 & IR source \\

144 & 2MASS J17595633-2403563$^{1}$& 2.81 & IR source \\

145 & 2MASS J17595847-2403304$^{1}$& 3.23 & IR source \\

146 & 2MASS J18005565-2403020$^{1}$ & 2.10 & IR source \\

147 & 2MASS J18012609-2401099$^{1}$& 1.05 & IR source \\

148 & 2MASS J18004417-2357518$^{1}$ & 1.46 & IR source \\

149 & 2MASS J17595956-2356255$^{1}$ & 2.67 & IR source \\

150 & 2MASS J18002106-2356145$^{1}$ & 3.20 & IR source \\

151 & 2MASS J18000452-2355587$^{1}$ & 0.94 & IR source \\

152 & 2MASS J18000800-2355319$^{1}$ & 1.24 & IR source \\

153 & 2MASS J18001137-2354403 $^{1}$ & 3.25 & IR source \\

155 & 2MASS J18011521-2410146$^{1}$ & 4.56 & IR source \\

156 & 2MASS J18005936-2404467$^{1}$ & 3.55 & IR source \\

157 & 2MASS J17595345-2404439$^{1}$ & 0.66 & IR source \\

159 & 2MASS J18005257-2404133$^{1}$ & 0.26 & IR source \\

\hline\end{tabular}
\caption{(continuation of Table B8) Possible counterparts from SIMBAD or 2MASS point source catalogue within $2*\rm{PSF}$ of numbered X-ray sources. 
  Correlation with non point like objects have been excluded. $^{\dagger}$ - Distance between centroid of X-ray source identified by \texttt{wavdetect} 
  and central position of possible counterpart as defined in the SIMBAD database or 2MASS point source catalogue. $^{A}$- identified within G5.89-0.39A, $^{B}$- identified within G5.89-0.39B.
  $^{1}$ - \citet{Braz:1983fj} Catalogue of non stellar molecular maser sources and their probable infrared counterparts.
  $^{2}$ - \citet{Cutri:2003kx} 2MASS All Sky Catalog of point sources. 
  $^{3}$ - \citet{Benjamin:2003yq} GLIMPSE. I. An SIRTF legacy project to map the inner galaxy.
  $^{4}$ - \citet{Hunter:2008fr} Subarcsecond submillimeter imaging of the ultracompact H II region G5.89-0.39.}
\label{tab:counter3}
\end{table*}

\section{K$\alpha$ emission line evidence in background regions}
\begin{figure*}
  \centering
 \includegraphics[width=0.95\textwidth]{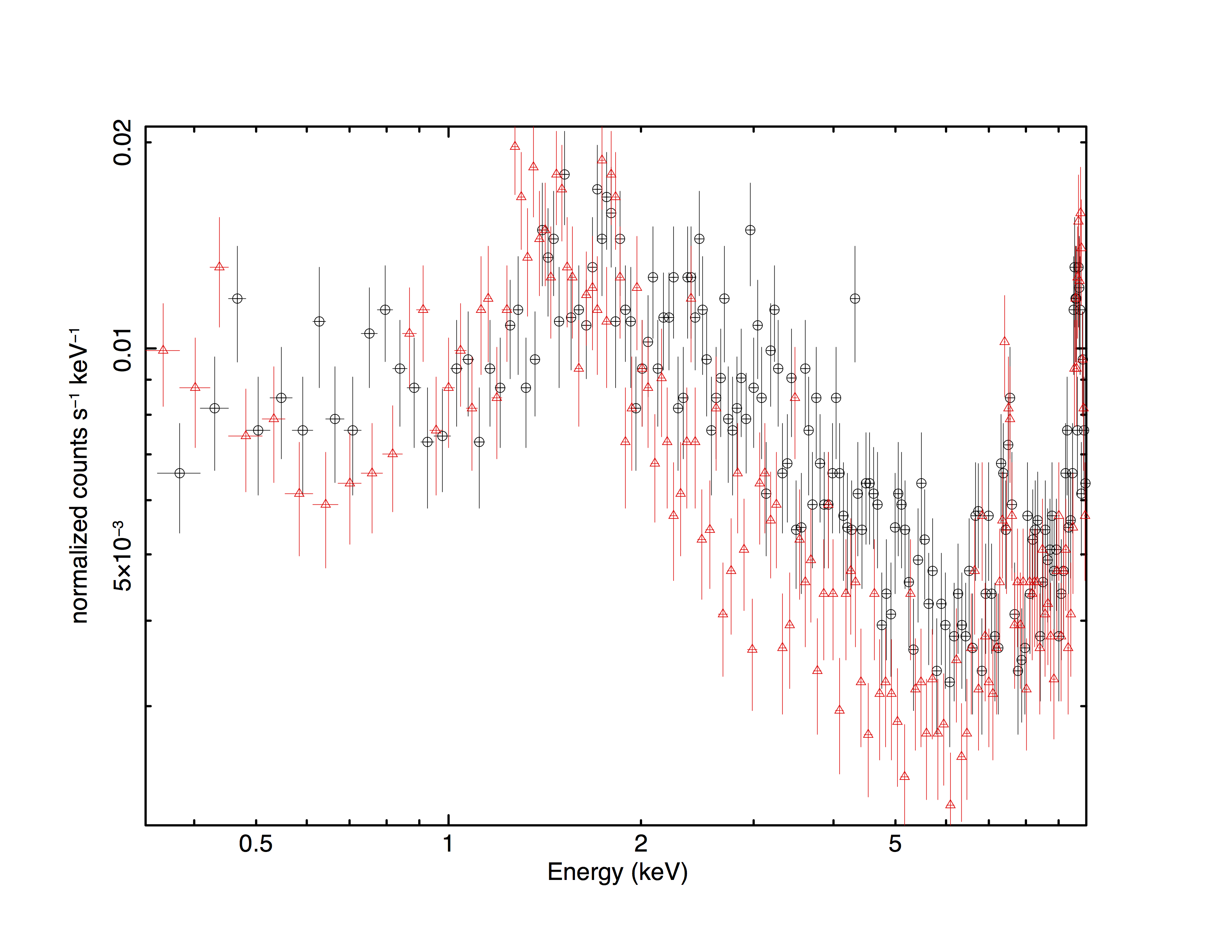}
  \caption{Black points: Energy spectrum of G5.89-0.39A+B with no background subtraction, minus the 35 identified sources. Red points: Energy spectrum of the background regions as indicated in Figure \ref{fig:bkgs}. Both spectra are normalised by ratios of their angular area, effective area, and exposure times.}
  \label{fig:kalph}
\end{figure*}

Figure \ref{fig:kalph} presents the G5.89-0.39A+B combined residual energy spectrum without background subtraction, and the raw energy spectrum of the background regions as shown in Figure \ref{fig:bkgs}. Both spectra reveal the 9.7 keV fluorescence line attributed to Au L$\alpha$ generated within the telescope to which can become dominant for high background to source area ratios for extended regions such as this. The spectra also show the presence of a peak around 6.7-7\,keV where we would expect the K$\alpha$ emission line to be present. The emission line is present in both the source and background spectra indicating that it is most likely due to the GDXE and not a localised source in the HII complex.

\section{Background regions for Unresolved sources}
\begin{figure*}
  \centering
 \includegraphics[width=0.95\textwidth]{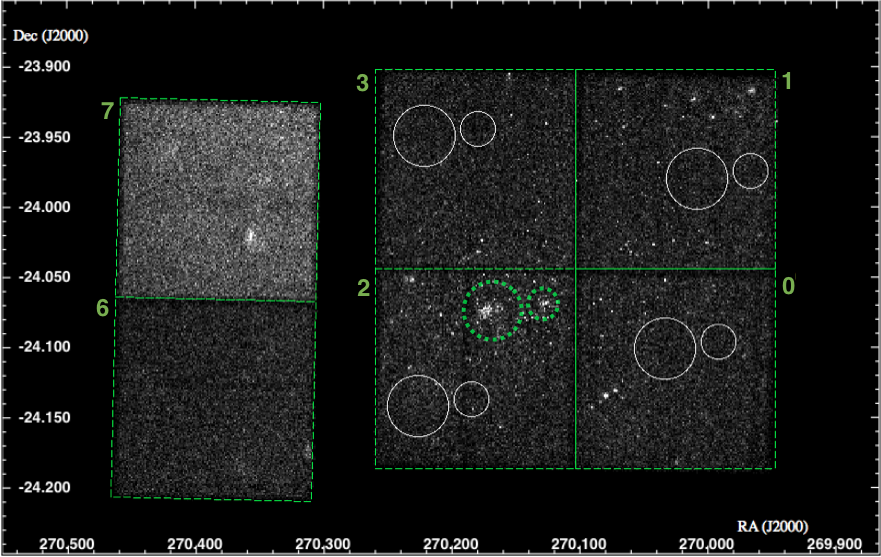}
  \caption{FoV of Chandra observation. White circles indicate the regions extracted as background regions. Green dashed squares indicate the Chandra CCDs. Green dashed circles indicate the HII regions G5.89-0.39A and B.}
  \label{fig:bkgs}
\end{figure*}

Figure \ref{fig:bkgs} indicates the regions extracted to obtain an average background count for the calculation of possible unresolved sources in G5.89-0.39, see \S\ref{sec:extended}.

\clearpage
\section{Table of Source Data}
\label{ap:table}
Description of each column present in Table \ref{tab:lums}: Table of Sources and Their Values.
\begin{enumerate}
\item Source \# - identifier used for each source identified through \texttt{wavdetect} and determined to be a significant source. Listed in order detected.

\item RA - Right Ascension position of centre of source identified, in degrees (J2000).
\item Dec - Declination position of centre of source identified, in degrees (J2000).
\item Net Counts 0.3-10 keV - The net number of photon events in the full energy range within the region defined for the source.
\item Net Counts 0.3-2.5 keV - The net number of photon events in the soft energy range within the region defined for the source.
\item Net Counts 2.5-5 keV - The net number of photon events in the medium energy range within the region defined for the source.
\item Net Counts 5-10 keV - The net number of photon events in the hard energy range within the region defined for the source.
\item Photon Flux [photons/$\rm{\, cm^{-2}\,s^{-1}}$] - Calculated over 0.3-10 keV energy range.
\item Energy Flux [$\rm{erg\, cm^{-2}\,s^{-1}}$] - Calculated over 0.3-10 keV energy range.
\item Observed Luminosity [erg $\rm{s^{-1}}$] - Observed luminosity as defined in \S\ref{sec:lumdeabs}. Calculated over 0.3-10 keV energy range.
\item Un-absorbed Luminosity [erg $s^{-1}$] - As defined in \S\ref{sec:lumdeabs}. Calculated over 0.3-10 keV energy range.  Only sources identified within the HII complex. 
\item Chandra ID - Chandra Discovered Source Name. CXO (Chandra X-ray Observatory)
\item Source Variability Index - 0-2 not variable, 2-5 possibly variable, >6 definitely variable.
\end{enumerate}

\begin{landscape}
\begin{table}
\caption{Table of Identified Sources and Their Values}
\setlength{\tabcolsep}{0pt}
\centering
\begin{tabular}{cccccccccccc}
\hline
 \multicolumn{1}{c}{} &
  \multicolumn{1}{c}{} &
  \multicolumn{1}{c}{} &
  \multicolumn{4}{c}{} &
  \multicolumn{2}{c}{Fluxes$^{\ddagger}$}&
  \multicolumn{2}{c}{Luminosities$^{\dagger}$} &
  \multicolumn{1}{c}{} 
   \\
  
  \multicolumn{1}{c}{Source \#} &
  \multicolumn{1}{c}{RA} &
  \multicolumn{1}{c}{Dec} &
  \multicolumn{4}{c}{Net Counts} &
  \multicolumn{1}{c}{$\rm{log \, F_{P}}$}&
  \multicolumn{1}{c}{$\rm{log \, F_{E}}$} &
  \multicolumn{1}{c}{$\rm{log\, L_{f}}$} &
  \multicolumn{1}{c}{$\rm{log \, L_{f,c}}$} &
  \multicolumn{1}{c}{Chandra ID} \\
  &
   \multicolumn{1}{c}{degrees} & 
   \multicolumn{1}{c}{degrees} &
  \multicolumn{1}{c}{0.3-10 keV} &
  \multicolumn{1}{c}{0.3-2.5 keV} &
  \multicolumn{1}{c}{2.5-5 keV} &
  \multicolumn{1}{c}{5-10 keV} &
  \multicolumn{1}{c}{($\rm{ph\,cm^{-2}\,s^{-1}}$)}&
  \multicolumn{1}{c}{($\rm{erg\,cm^{-2}\,s^{-1}}$)}&
  \multicolumn{1}{c}{($\rm{erg\,s^{-1}}$)}&
  \multicolumn{1}{c}{($\rm{erg\,s^{-1}}$)}&
  \\
  \multicolumn{1}{c}{(1)}&
  \multicolumn{1}{c}{(2)}&
  \multicolumn{1}{c}{(3)}&
  \multicolumn{1}{c}{(4)}&
  \multicolumn{1}{c}{(5)}&
  \multicolumn{1}{c}{(6)}&
  \multicolumn{1}{c}{(7)}&
  \multicolumn{1}{c}{(8)}&
  \multicolumn{1}{c}{(9)}&
  \multicolumn{1}{c}{(10)}&
  \multicolumn{1}{c}{(11)}&
  \multicolumn{1}{c}{(12)}\\
\hline \\
1 & 270.08 & -24.134 & $569.68 \pm 25.06$ & $40.1 \pm 6.86$ & $346.88 \pm 19.11$ & $179.16 \pm 14.39$ & -4.39 & -12.48 & 32.2 & - & CXOU J180019.2-240801\\
2 & 270.142 & -24.098 & $41.32 \pm 6.86$ & $11.01 \pm 3.46$ & $27.65 \pm 5.39$ & - & -5.73 & -13.9 & 30.78 & - & CXOU J180034.1-240553\\
3 & 270.144 & -24.077 & $33.52 \pm 6.16$ & $10.65 \pm 3.32$ & $15.96 \pm 4.12$ & - & -5.52 & -13.46 & 31.22 & - & CXOU J180034.1-240439\\
4$\rm{^{B}}$ & 270.129 & -24.077 & $35.3 \pm 6.16$ & $9.56 \pm 3.16$ & $15.59 \pm 4.0$ & - & -5.78 & -13.94 & 30.74 & 31.07 & CXOU J180030.9-240438\\
5$\rm{^{B}}$ & 270.144 & -24.074 & $40.48 \pm 6.71$ & $5.8 \pm 2.45$ & $12.65 \pm 3.61$ & - & -5.91 & -14.19 & 30.49 & 30.82 & CXOU J180034.6-240426\\
6$\rm{^{B}}$ & 270.128 & -24.071 & $25.43 \pm 5.2$ & - & $18.34 \pm 4.36$ & $6.59 \pm 2.65$ & -5.89 & -13.97 & 30.71 & 31.04 & CXOU J180030.7-240417\\
7$\rm{^{B}}$ & 270.124 & -24.069 & $15.58 \pm 4.12$ & $6.65 \pm 2.65$ & $9.67 \pm 3.16$ & - & -6.29 & -14.56 & 30.12 & 30.45 & CXOU J180029.7-240410\\
8$\rm{^{A}}$ & 270.171 & -24.069 & $80.81 \pm 9.64$ & $14.07 \pm 4.0$ & $52.64 \pm 7.55$ & $11.65 \pm 4.0$ & -5.23 & -13.65 & 31.03 & 31.35 & CXOU J180041.1-240409\\
9$\rm{^{B}}$ & 270.131 & -24.068 & $31.36 \pm 5.83$ & $9.48 \pm 3.16$ & $12.46 \pm 3.61$ & - & -6.01 & -14.24 & 30.44 & 30.76 & CXOU J180031.5-240405\\
10$\rm{^{B}}$ & 270.129 & -24.068 & $16.69 \pm 4.36$ & - & $11.42 \pm 3.46$ & - & -5.78 & -13.7 & 30.98 & 31.31 & CXOU J180030.9-240405\\
11 & 270.089 & -24.067 & $81.06 \pm 9.11$ & $21.53 \pm 4.69$ & $48.57 \pm 7.0$ & $12.63 \pm 3.74$ & -5.51 & -13.68 & 31.0 & - & CXOU J180021.4-240402\\
12$\rm{^{B}}$ & 270.118 & -24.066 & $18.44 \pm 4.47$ & - & $8.8 \pm 3.0$ & $7.59 \pm 2.83$ & -5.82 & -14.02 & 30.66 & 30.98 & CXOU J180028.3-240358\\
13$\rm{^{B}}$ & 270.127 & -24.062 & $15.95 \pm 4.12$ & - & $11.63 \pm 3.46$ & - & -6.25 & -14.47 & 30.21 & 30.54 & CXOU J180030.5-240342\\
14 & 270.115 & -24.059 & $10.16 \pm 3.32$ & - & $7.82 \pm 2.83$ & - & -6.52 & -14.7 & 29.98 & - & CXOU J180027.5-240334 \\
15$\rm{^{A}}$ & 270.168 & -24.052 & $19.12 \pm 4.69$ & $11.02 \pm 3.46$ & $8.46 \pm 3.0$ & - & -6.18 & -14.56 & 30.12 & 30.45 & CXOU J180040.3-240308\\
16 & 270.136 & -24.052 & $97.54 \pm 10.0$ & $17.66 \pm 4.24$ & $48.33 \pm 7.0$ & $26.53 \pm 5.2$ & -5.34 & -13.41 & 31.27 & - & CXOU J180032.7-240306\\
17 & 270.121 & -24.05 & $14.7 \pm 4.0$ & $13.59 \pm 3.74$ & - & - & -6.38 & -14.91 & 29.77 & - & CXOU J180029.1-240300\\
18 & 270.143 & -24.049 & $15.86 \pm 4.12$ & $7.68 \pm 2.83$ & $6.69 \pm 2.65$ & - & -6.19 & -14.4 & 30.28 & - & CXOU J180034.3-240258\\
19 & 270.156 & -24.048 & $89.4 \pm 9.59$ & $40.32 \pm 6.4$ & $40.32 \pm 6.4$ & - & -5.57 & -13.87 & 30.81 & - & CXOU J180037.6-240251\\
20 & 270.056 & -24.032 & $26.85 \pm 5.48$ & $26.67 \pm 5.29$ & - & - & -5.9 & -14.35 & 30.33 & - & CXOU J180013.5-240156\\

\\
\hline\end{tabular}
\\
\begin{flushleft} 
$\rm{^{A}}$ - Source identified in the conventional HII region, G5.89-0.39A\\
$\rm{^{B}}$ - Source identified in the UC HII region, G5.89-0.39B\\
$^{*}$ - Feldt's star (see ~\S\ref{sec:feldt})\\
$\rm{^{\ddagger}}$ - X-ray fluxes: Calculated for the full energy band (0.3-10.0 keV); P - photon flux; E - energy flux. Average error in fluxes $\sim 25 \%$. Errors calculated from error in net counts.\\
$\rm{^{\dagger}}$ - X-ray luminosities: f = full band (0.3-10.0 keV). Absorption corrected luminosity subscripted with a c. Luminosities were calculated assuming a distance of 2 kpc. Average error in the luminosity is $\sim 25 \%$. Errors calculated from error in net counts.
\end{flushleft}
\label{tab:lums}
\end{table}
\end{landscape}
\begin{landscape}
\begin{table}
\caption{Table of Identified Sources and Their Values}
\setlength{\tabcolsep}{0pt}
\centering
\begin{tabular}{cccccccccccc}
\hline
 \multicolumn{1}{c}{} &
  \multicolumn{1}{c}{} &
  \multicolumn{1}{c}{} &
  \multicolumn{4}{c}{} &
  \multicolumn{2}{c}{Fluxes$^{\ddagger}$}&
  \multicolumn{2}{c}{Luminosities$^{\dagger}$} &
   \multicolumn{1}{c}{}
   \\
  
  \multicolumn{1}{c}{Source \#} &
  \multicolumn{1}{c}{RA} &
  \multicolumn{1}{c}{Dec} &
  \multicolumn{4}{c}{Net Counts} &
  \multicolumn{1}{c}{$\rm{log \, F_{P}}$}&
  \multicolumn{1}{c}{$\rm{log \, F_{E}}$} &
  \multicolumn{1}{c}{$\rm{log\, L_{f}}$} &
  \multicolumn{1}{c}{$\rm{log \, L_{f,c}}$} &
   \multicolumn{1}{c}{Chandra ID}\\
  &
   \multicolumn{1}{c}{degrees} & 
   \multicolumn{1}{c}{degrees} &
  \multicolumn{1}{c}{0.3-10 keV} &
  \multicolumn{1}{c}{0.3-2.5 keV} &
  \multicolumn{1}{c}{2.5-5 keV} &
  \multicolumn{1}{c}{5-10 keV} &
  \multicolumn{1}{c}{($\rm{ph\,cm^{-2}\,s^{-1}}$)}&
  \multicolumn{1}{c}{($\rm{erg\,cm^{-2}\,s^{-1}}$)}&
  \multicolumn{1}{c}{($\rm{erg\,s^{-1}}$)}&
  \multicolumn{1}{c}{($\rm{erg\,s^{-1}}$)}&
   \multicolumn{1}{c}{}
  \\
  \multicolumn{1}{c}{(1)}&
  \multicolumn{1}{c}{(2)}&
  \multicolumn{1}{c}{(3)}&
  \multicolumn{1}{c}{(4)}&
  \multicolumn{1}{c}{(5)}&
  \multicolumn{1}{c}{(6)}&
  \multicolumn{1}{c}{(7)}&
  \multicolumn{1}{c}{(8)}&
  \multicolumn{1}{c}{(9)}&
  \multicolumn{1}{c}{(10)}&
  \multicolumn{1}{c}{(11)} &
   \multicolumn{1}{c}{12}\\
\hline \\
21 & 270.18 & -24.03 & $434.94 \pm 21.12$ & $334.53 \pm 18.41$ & $83.63 \pm 9.22$ & $13.47 \pm 4.12$ & -4.79 & -13.21 & 31.47 & - & CXOU J180043.3-240150\\
22 & 270.128 & -24.026 & $55.41 \pm 7.55$ & $10.68 \pm 3.32$ & $35.59 \pm 6.0$ & $8.59 \pm 3.0$ & -5.68 & -13.92 & 30.76 & - & CXOU J180030.8-240133\\
23 & 270.042 & -24.025 & $178.87 \pm 13.96$ & $59.23 \pm 7.81$ & $92.59 \pm 9.8$ & $24.93 \pm 5.57$ & -5.09 & -13.24 & 31.44 & - & CXOU J180010.0-240129\\
24 & 270.126 & -24.023 & $32.61 \pm 5.83$ & $13.61 \pm 3.74$ & $14.77 \pm 3.87$ & - & -5.99 & -14.3 & 30.38 & - & CXOU J180030.3-240123\\
25 & 270.177 & -24.022 & $120.9 \pm 11.31$ & $38.25 \pm 6.32$ & $62.58 \pm 8.0$ & $14.94 \pm 4.24$ & -5.24 & -13.37 & 31.31 & - & CXOU J180042.5-240120\\
26 & 270.141 & -24.015 & $26.84 \pm 5.29$ & - & $12.77 \pm 3.61$ & $15.32 \pm 4.12$ & -5.8 & -13.79 & 30.89 & - & CXOU J180033.9-240054\\
27 & 270.07 & -24.008 & $31.41 \pm 5.83$ & $26.18 \pm 5.2$ & - & - & -4.09 & -13.34 & 31.34 & - & CXOU J180016.9-240030\\
28 & 270.135 & -24.008 & $36.3 \pm 6.32$ & - & $15.74 \pm 4.0$ & $16.24 \pm 4.24$ & -5.49 & -13.45 & 31.23 & - & CXOU J180032.3-240027\\
29 & 270.13 & -24.002 & $10.79 \pm 3.61$ & $5.77 \pm 2.45$ & - & - & -6.21 & -14.4 & 30.28 & - & CXOU J180031.3-240006\\
30 & 270.133 & -23.994 & $45.0 \pm 7.0$ & $34.04 \pm 5.92$ & $9.52 \pm 3.16$ & - & -5.92 & -14.41 & 30.27 & - & CXOU J180031.9-235940\\
31 & 270.091 & -24.142 & $32.99 \pm 8.49$ & - & $14.51 \pm 4.8$ & - & -4.63 & -12.7 & 31.98 & -  & CXOU J180021.8-240831\\
32 & 270.072 & -24.13 & $107.53 \pm 11.79$ & - & $60.83 \pm 8.25$ & $41.11 \pm 7.87$ & -4.93 & -12.86 & 31.82 & - &CXOU J180017.4-240747\\
33 & 270.136 & -24.102 & $13.9 \pm 4.12$ & $12.67 \pm 3.74$ & - & - & -5.82 & -14.29 & 30.39 & - & CXOU J180032.7-240608\\
34 & 270.106 & -24.1 & $11.6 \pm 3.61$ & $8.59 \pm 3.0$ & - & - & -6.15 & -14.37 & 30.31 & - & CXOU J180025.4-240559\\
35 & 270.123 & -24.096 & $23.24 \pm 5.1$ & $11.25 \pm 3.46$ & $9.4 \pm 3.16$ & - & -5.95 & -14.11 & 30.57 & - & CXOU J180029.6-240545\\
36$\rm{^{A}}$ & 270.164 & -24.092 & $40.74 \pm 7.75$ & $9.33 \pm 3.32$ & $14.72 \pm 4.12$ & - & -5.61 & -13.7 & 30.98 & 31.3 & CXOU J180039.4-240531\\
37 & 270.129 & -24.092 & $12.73 \pm 3.87$ & - & $6.54 \pm 2.65$ & - & -6.24 & -14.56 & 30.12 & - & CXOU J180031.0-240531\\
38 & 270.083 & -24.09 & $22.8 \pm 5.1$ & $21.58 \pm 4.8$ & - & - & -4.92 & -14.04 & 30.64 & - & CXOU J180019.9-240526\\
39 & 270.2 & -24.089 & $71.37 \pm 9.7$ & $29.81 \pm 6.0$ & $27.35 \pm 5.57$ & - & -5.19 & -13.16 & 31.52 & - & CXOU J180048.0-240520\\
40 & 270.152 & -24.088 & $15.78 \pm 4.47$ & - & $7.29 \pm 2.83$ & - & -5.7 & -13.62 & 31.06 & - & CXOU J180036.5-240518\\

\\
\hline\end{tabular}
\\
\begin{flushleft} 
  $\rm{^{A}}$ - Source identified in the conventional HII region, G5.89-0.39A\\
  $\rm{^{B}}$ - Source identified in the UC HII region, G5.89-0.39B\\
  $\rm{^{\ddagger}}$ - X-ray fluxes: Calculated for the full energy band (0.3-10.0 keV); P - photon flux; E - energy flux. 
  Average error in fluxes $\sim 25 \%$. Errors calculated from error in net counts.\\
  $\rm{^{\dagger}}$ - X-ray luminosities: f = full band (0.3-10.0 keV). Absorption corrected luminosity subscripted with a c. Luminosities were calculated assuming 
  a distance of 2 kpc. Average error in the luminosity is $\sim 25 \%$. Errors calculated from error in net counts.
\end{flushleft}
\end{table}
\end{landscape}

\begin{landscape}
\begin{table}
\caption{table of Identified Sources and Their Values}
\setlength{\tabcolsep}{0pt}
\centering
\begin{tabular}{cccccccccccc}
\hline
 \multicolumn{1}{c}{} &
  \multicolumn{1}{c}{} &
  \multicolumn{1}{c}{} &
  \multicolumn{4}{c}{} &
  \multicolumn{2}{c}{Fluxes$^{\ddagger}$}&
  \multicolumn{2}{c}{Luminosities$^{\dagger}$} &
  \multicolumn{1}{c}{}
   \\
  
  \multicolumn{1}{c}{Source \#} &
  \multicolumn{1}{c}{RA} &
  \multicolumn{1}{c}{Dec} &
  \multicolumn{4}{c}{Net Counts} &
  \multicolumn{1}{c}{$\rm{log \, F_{P}}$}&
  \multicolumn{1}{c}{$\rm{log \, F_{E}}$} &
  \multicolumn{1}{c}{$\rm{log\, L_{f}}$} &
  \multicolumn{1}{c}{$\rm{log \, L_{f,c}}$} &
  \multicolumn{1}{c}{Chandra ID}\\
  &
   \multicolumn{1}{c}{degrees} & 
   \multicolumn{1}{c}{degrees} &
  \multicolumn{1}{c}{0.3-10 keV} &
  \multicolumn{1}{c}{0.3-2.5 keV} &
  \multicolumn{1}{c}{2.5-5 keV} &
  \multicolumn{1}{c}{5-10 keV} &
  \multicolumn{1}{c}{($\rm{ph\,cm^{-2}\,s^{-1}}$)}&
  \multicolumn{1}{c}{($\rm{erg\,cm^{-2}\,s^{-1}}$)}&
  \multicolumn{1}{c}{($\rm{erg\,s^{-1}}$)}&
  \multicolumn{1}{c}{($\rm{erg\,s^{-1}}$)}&
  \multicolumn{1}{c}{}
  \\
  \multicolumn{1}{c}{(1)}&
  \multicolumn{1}{c}{(2)}&
  \multicolumn{1}{c}{(3)}&
  \multicolumn{1}{c}{(4)}&
  \multicolumn{1}{c}{(5)}&
  \multicolumn{1}{c}{(6)}&
  \multicolumn{1}{c}{(7)}&
  \multicolumn{1}{c}{(8)}&
  \multicolumn{1}{c}{(9)}&
  \multicolumn{1}{c}{(10)}&
  \multicolumn{1}{c}{(11)}&
  \multicolumn{1}{c}{(12)}\\
\hline \\

41 & 270.115 & -24.088 & $14.75 \pm 4.24$ & - & $8.4 \pm 3.0$ & - & -5.55 & -13.42 & 31.26 & - & CXOU J180027.7-240517\\
42 & 270.108 & -24.085 & $58.03 \pm 7.87$ & $10.52 \pm 3.32$ & $38.28 \pm 6.25$ & $8.62 \pm 3.16$ & -5.62 & -13.79 & 30.89 & - & CXOU J180026.0-240506\\
43$\rm{^{A}}$ & 270.164 & -24.085 & $30.72 \pm 6.16$ & - & $16.99 \pm 4.36$ & - & -5.61 & -13.63 & 31.05 & 31.38 & CXOU J180039.4-240506\\
44$\rm{^{A}}$ & 270.178 & -24.085 & $29.19 \pm 6.25$ & $9.57 \pm 3.46$ & $13.56 \pm 4.0$ & - & -5.3 & -13.25 & 31.43 & 31.75 & CXOU J180042.8-240505\\
45 & 270.117 & -24.082 & $24.97 \pm 5.2$ & - & $13.54 \pm 3.74$ & $10.72 \pm 3.46$ & -5.83 & -13.85 & 30.83 & - & CXOU J180028.1-240455\\
46$\rm{^{A}}$ & 270.17 & -24.082 & $21.63 \pm 5.1$ & $7.65 \pm 3.0$ & $9.91 \pm 3.32$ & - & -5.65 & -13.57 & 31.11 & 31.43 & CXOU J180040.9-240454\\
47 & 270.142 & -24.079 & $22.9 \pm 5.2$ & $4.71 \pm 2.24$ & $11.85 \pm 3.61$ & - & -5.84 & -13.97 & 30.71 & - & CXOU J180034.0-240446\\
48 & 270.15 & -24.079 & $29.11 \pm 5.74$ & $25.45 \pm 5.2$ & - & - & -5.99 & -14.48 & 30.2 & - & CXOU J180036.0-240443\\
49$\rm{^{B}}$ & 270.125 & -24.077 & $10.05 \pm 3.32$ & $3.8 \pm 2.0$ & $5.71 \pm 2.45$ & - & -6.47 & -14.79 & 29.89 & 30.22 & CXOU J180030.0-240439\\
50$\rm{^{B}}$ & 270.137 & -24.077 & $17.06 \pm 4.47$ & $8.68 \pm 3.0$ & - & - & -6.2 & -14.48 & 30.2 & 30.52 & CXOU J180032.9-240438\\
51$\rm{^{A}}$ & 270.175 & -24.077 & $43.11 \pm 8.25$ & - & $21.33 \pm 5.0$ & -& -5.08 & -13.04 & 31.64 & 31.96 & CXOU J180041.9-240436\\
52 & 270.049 & -24.075 & $55.99 \pm 8.37$ & $13.87 \pm 3.87$ & $25.46 \pm 5.2$ & $13.01 \pm 4.24$ & -5.05 & -12.96 & 31.72 & - & CXOU J180011.8-240428\\
53$\rm{^{A}}$ & 270.176 & -24.074 & $28.67 \pm 6.25$ & $15.81 \pm 4.36$ & $12.18 \pm 3.87$ & - & -5.63 & -13.94 & 30.74 & 31.06 & CXOU J180042.2-240426\\
54 & 270.075 & -24.073 & $33.31 \pm 6.16$ & - & $12.31 \pm 3.61$ & $19.86 \pm 4.69$ & -5.43 & -13.4 & 31.28 & - & CXOU J180018.1-240422\\
55$\rm{^{A}}$ & 270.177 & -24.072 & $37.24 \pm 6.86$ & $13.61 \pm 4.0$ & $17.12 \pm 4.47$ & - & -5.48 & -13.48 & 31.2 & 31.53 & CXOU J180042.4-240420\\
56$\rm{^{A}}$ & 270.179 & -24.072 & $44.6 \pm 8.0$ & $11.97 \pm 3.74$ & $13.1 \pm 3.87$ & -& -5.6 & -13.61 & 31.07 & 31.4 & CXOU J180043.0-240420\\
57$\rm{^{A}}$ & 270.161 & -24.071 & $24.68 \pm 5.57$ & $11.04 \pm 3.61$ & $14.38 \pm 4.0$ & - & -5.54 & -13.47 & 31.21 & 31.54 & CXOU J180038.7-240417\\
58$\rm{^{A}}$ & 270.17 & -24.071 & $24.37 \pm 5.66$ & $9.12 \pm 3.32$ & $10.61 \pm 3.61$ & - & -4.94 & -13.54 & 31.14 & 31.46 & CXOU J180040.8-240415\\
59$\rm{^{A}}$ & 270.176 & -24.071 & $26.87 \pm 5.83$ & - & $14.02 \pm 4.0$ & - & -5.77 & -13.82 & 30.86 & 31.19 & CXOU J180042.2-240415\\
60$\rm{^{A}}$ & 270.164 & -24.07 & $21.53 \pm 5.2$ & $7.64 \pm 3.0$ & - & - & -5.65 & -13.6 & 31.08 & 31.4 & CXOU J180039.5-240411\\

\\
\hline\end{tabular}
\\
\begin{flushleft} 
  $\rm{^{A}}$ - Source identified in the conventional HII region, G5.89-0.39A\\
  $\rm{^{B}}$ - Source identified in the UC HII region, G5.89-0.39B\\
  $\rm{^{\ddagger}}$ - X-ray fluxes: Calculated for the full energy band (0.3-10.0 keV); P - photon flux; E - energy flux. 
  Average error in fluxes $\sim 25 \%$. Errors calculated from error in net counts.\\
  $\rm{^{\dagger}}$ - X-ray luminosities: f = full band (0.3-10.0 keV). Absorption corrected luminosity subscripted with a c. Luminosities were calculated assuming 
  a distance of 2 kpc. Average error in the luminosity is $\sim 25 \%$. Errors calculated from error in net counts.
\end{flushleft}
\end{table}
\end{landscape}

\begin{landscape}
\begin{table}
\caption{Table of Identified Sources and Their Values}
\setlength{\tabcolsep}{0pt}
\centering
\begin{tabular}{cccccccccccc}
\hline
 \multicolumn{1}{c}{} &
  \multicolumn{1}{c}{} &
  \multicolumn{1}{c}{} &
  \multicolumn{4}{c}{} &
  \multicolumn{2}{c}{Fluxes$^{\ddagger}$}&
  \multicolumn{2}{c}{Luminosities$^{\dagger}$} &
  \multicolumn{1}{c}{}
   \\
  
  \multicolumn{1}{c}{Source \#} &
  \multicolumn{1}{c}{RA} &
  \multicolumn{1}{c}{Dec} &
  \multicolumn{4}{c}{Net Counts} &
  \multicolumn{1}{c}{$\rm{log \, F_{P}}$}&
  \multicolumn{1}{c}{$\rm{log \, F_{E}}$} &
  \multicolumn{1}{c}{$\rm{log\, L_{f}}$} &
  \multicolumn{1}{c}{$\rm{log \, L_{f,c}}$} &
  \multicolumn{1}{c}{Chandra ID}\\
  &
   \multicolumn{1}{c}{degrees} & 
   \multicolumn{1}{c}{degrees} &
  \multicolumn{1}{c}{0.3-10 keV} &
  \multicolumn{1}{c}{0.3-2.5 keV} &
  \multicolumn{1}{c}{2.5-5 keV} &
  \multicolumn{1}{c}{5-10 keV} &
  \multicolumn{1}{c}{($\rm{ph\,cm^{-2}\,s^{-1}}$)}&
  \multicolumn{1}{c}{($\rm{erg\,cm^{-2}\,s^{-1}}$)}&
  \multicolumn{1}{c}{($\rm{erg\,s^{-1}}$)}&
  \multicolumn{1}{c}{($\rm{erg\,s^{-1}}$)}&
  \multicolumn{1}{c}{}
  \\
  \multicolumn{1}{c}{(1)}&
  \multicolumn{1}{c}{(2)}&
  \multicolumn{1}{c}{(3)}&
  \multicolumn{1}{c}{(4)}&
  \multicolumn{1}{c}{(5)}&
  \multicolumn{1}{c}{(6)}&
  \multicolumn{1}{c}{(7)}&
  \multicolumn{1}{c}{(8)}&
  \multicolumn{1}{c}{(9)}&
  \multicolumn{1}{c}{(10)}&
  \multicolumn{1}{c}{(11)}&
  \multicolumn{1}{c}{(12)}\\
\hline \\
61$\rm{^{A}}$ & 270.181 & -24.068 & $22.33 \pm 5.66$ & $8.94 \pm 3.16$ & $8.52 \pm 3.16$ & - & -5.44 & -13.36 & 31.32 & 31.65 & CXOU J180043.4-240403\\
62$\rm{^{B}}$ & 270.126 & -24.067 & $27.79 \pm 5.48$ & - & $25.93 \pm 5.2$ & $3.72 \pm 2.0$ & -5.83 & -13.96 & 30.72 & 31.04 & CXOU J180030.3-240401\\
63 & 270.05 & -24.066 & $12.48 \pm 3.87$ & $8.06 \pm 3.0$ & - & - & -6.26 & -14.72 & 29.96 & - & CXOU J180012.1-240359\\
64 & 270.076 & -24.065 & $10.16 \pm 3.46$ & $6.67 \pm 2.65$ & - & - & -6.43 & -14.8 & 29.88 & - & CXOU J180018.3-240353\\
65$\rm{^{A}}$ & 270.155 & -24.065 & $17.12 \pm 4.58$ & - & $10.36 \pm 3.32$ & - & -4.98 & -13.96 & 30.72 & 31.05 & CXOU J180037.2-240352\\
66$\rm{^{B}}$ & 270.133 & -24.064 & $10.46 \pm 3.46$ & - & $5.64 \pm 2.45$ & - & -6.45 & -14.76 & 29.92 & 30.25 & CXOU J180031.9-240351\\
67 & 270.084 & -24.064 & $14.88 \pm 4.0$ & $15.43 \pm 4.0$ & $5.78 \pm 2.45$ & $12.46 \pm 3.74$ & -5.72 & -13.97 & 30.71 & - & CXOU J180020.0-240351\\
68$\rm{^{A}}$ & 270.167 & -24.064 & $38.39 \pm 6.56$ & $15.59 \pm 4.12$ & $14.82 \pm 4.0$ & $6.58 \pm 2.83$ & -5.44 & -13.38 & 31.3 & 31.63 & CXOU J180040.1-240351\\
69$\rm{^{A}}$ & 270.18 & -24.064 & $14.85 \pm 4.24$ & - & $6.38 \pm 2.65$ & - & -5.65 & -13.61 & 31.07 & 31.4 & CXOU J180043.3-240350\\
70$\rm{^{B}}$ & 270.131 & -24.063 & $17.59 \pm 4.36$ & -& $13.49 \pm 3.74$ & - & -6.17 & -14.5 & 30.18 & 30.5 & CXOU J180031.5-240345\\
71 & 270.053 & -24.059 & $11.49 \pm 3.74$ & $10.76 \pm 3.46$ & - & - & -6.3 & -14.9 & 29.78 & - & CXOU J180012.7-240331\\
72 & 270.196 & -24.056 & $14.2 \pm 4.8$ & - & $11.13 \pm 3.74$ & - & -5.99 & -14.4 & 30.28 & - & CXOU J180047.1-240322\\
73$\rm{^{A}}$ & 270.177 & -24.054 & $24.97 \pm 5.39$ & $8.07 \pm 3.0$ & $11.33 \pm 3.46$ & - & -5.77 & -13.92 & 30.76 & 31.08 & CXOU J180042.4-240316\\
74$\rm{^{A}}$ & 270.182 & -24.054 & $12.21 \pm 3.87$ & $8.07 \pm 3.0$ & - & - & -6.34 & -14.74 & 29.94 & 30.27 & CXOU J180043.7-240316\\
75 & 270.033 & -24.051 & $42.5 \pm 7.28$ & $37.5 \pm 6.4$ & - & - & -5.44 & -13.49 & 31.19 & - & CXOU J180007.9-240302\\
76 & 270.176 & -24.047 & $16.19 \pm 4.36$ & $12.73 \pm 3.74$ & - & - & -6.09 & -14.34 & 30.34 & - & CXOU J180042.1-240248\\
77 & 270.174 & -24.032 & $23.32 \pm 5.39$ & - & $16.76 \pm 4.24$ & - & -5.6 & -13.55 & 31.13 & - & CXOU J180041.7-240157\\
78 & 270.061 & -24.032 & $12.18 \pm 3.87$ & $11.93 \pm 3.61$ & - & - & -5.97 & -14.11 & 30.57 & - & CXOU J180014.7-240157\\
79 & 270.06 & -24.031 & $27.19 \pm 5.66$ & - & $16.92 \pm 4.24$ & - & -5.67 & -13.84 & 30.84 & - & CXOU J180014.4-240150\\
80 & 270.178 & -24.03 & $12.67 \pm 4.0$ & - & $4.62 \pm 2.24$ & - & -6.49 & -14.89 & 29.79 & - & CXOU J180042.8-240149\\

\\
\hline\end{tabular}
\\
\begin{flushleft} 
  $\rm{^{A}}$ - Source identified in the conventional HII region, G5.89-0.39A\\
  $\rm{^{B}}$ - Source identified in the UC HII region, G5.89-0.39B\\
  $\rm{^{\ddagger}}$ - X-ray fluxes: Calculated for the full energy band (0.3-10.0 keV); P - photon flux; E - energy flux. 
  Average error in fluxes $\sim 25 \%$. Errors calculated from error in net counts.\\
  $\rm{^{\dagger}}$ - X-ray luminosities: f = full band (0.3-10.0 keV). Absorption corrected luminosity subscripted with a c. Luminosities were 
  calculated assuming a distance of 2 kpc. Average error in the luminosity is $\sim 25 \%$. Errors calculated from error in net counts.
\end{flushleft}
\end{table}
\end{landscape}
\begin{landscape}
\begin{table}
\caption{Table of Identified Sources and Their Values}
\setlength{\tabcolsep}{0pt}
\centering
\begin{tabular}{cccccccccccc}
\hline
 \multicolumn{1}{c}{} &
  \multicolumn{1}{c}{} &
  \multicolumn{1}{c}{} &
  \multicolumn{4}{c}{} &
  \multicolumn{2}{c}{Fluxes$^{\ddagger}$}&
  \multicolumn{2}{c}{Luminosities$^{\dagger}$} &
  \multicolumn{1}{c}{}
   \\
  
  \multicolumn{1}{c}{Source \#} &
  \multicolumn{1}{c}{RA} &
  \multicolumn{1}{c}{Dec} &
  \multicolumn{4}{c}{Net Counts} &
  \multicolumn{1}{c}{$\rm{log \, F_{P}}$}&
  \multicolumn{1}{c}{$\rm{log \, F_{E}}$} &
  \multicolumn{1}{c}{$\rm{log\, L_{f}}$} &
  \multicolumn{1}{c}{$\rm{log \, L_{f,c}}$} &
  \multicolumn{1}{c}{Chandra ID} \\
  &
   \multicolumn{1}{c}{degrees} & 
   \multicolumn{1}{c}{degrees} &
  \multicolumn{1}{c}{0.3-10 keV} &
  \multicolumn{1}{c}{0.3-2.5 keV} &
  \multicolumn{1}{c}{2.5-5 keV} &
  \multicolumn{1}{c}{5-10 keV} &
  \multicolumn{1}{c}{($\rm{ph\,cm^{-2}\,s^{-1}}$)}&
  \multicolumn{1}{c}{($\rm{erg\,cm^{-2}\,s^{-1}}$)}&
  \multicolumn{1}{c}{($\rm{erg\,s^{-1}}$)}&
  \multicolumn{1}{c}{($\rm{erg\,s^{-1}}$)}&
  \multicolumn{1}{c}{}
  \\
  \multicolumn{1}{c}{(1)}&
  \multicolumn{1}{c}{(2)}&
  \multicolumn{1}{c}{(3)}&
  \multicolumn{1}{c}{(4)}&
  \multicolumn{1}{c}{(5)}&
  \multicolumn{1}{c}{(6)}&
  \multicolumn{1}{c}{(7)}&
  \multicolumn{1}{c}{(8)}&
  \multicolumn{1}{c}{(9)}&
  \multicolumn{1}{c}{(10)}&
  \multicolumn{1}{c}{(11)}&
  \multicolumn{1}{c}{(12)}\\
\hline \\
81 & 270.068 & -24.028 & $14.54 \pm 4.36$ & - & $9.31 \pm 3.16$ & - & -5.65 & -14.1 & 30.58 & - & CXOU J180016.3-240143\\
82 & 270.065 & -24.025 & $11.12 \pm 3.74$ & - & $5.49 \pm 2.45$ & - & -5.87 & -13.81 & 30.87 & - & CXOU J180015.5-240131\\
83 & 270.062 & -24.023 & $18.16 \pm 4.58$ & - & - & $14.28 \pm 4.0$ & -5.61 & -13.5 & 31.18 & - & CXOU J180014.8-240121\\
84 & 270.027 & -24.022 & $47.17 \pm 7.55$ & $40.11 \pm 6.56$ & $12.66 \pm 3.87$ & - & -5.35 & -14.0 & 30.68 & - & CXOU J180006.4-240120\\
85 & 270.179 & -24.019 & $10.9 \pm 3.61$ & - & $9.41 \pm 3.16$ & - & -6.12 & -14.18 & 30.5 & - & CXOU J180042.9-240108\\
86 & 270.035 & -23.999 & $15.57 \pm 4.58$ & $13.87 \pm 4.0$ & - & - & -5.71 & -13.76 & 30.92 & - & CXOU J180008.4-235957\\
87 & 270.048 & -23.99 & $22.25 \pm 5.57$ & $18.09 \pm 4.58$ & - & - & -5.47 & -13.4 & 31.28 & - & CXOU J180011.6-235924\\
88 & 270.073 & -23.981 & $11.93 \pm 4.24$ & $14.22 \pm 4.12$ & - & - & -5.77 & -14.33 & 30.35 & - & CXOU J180017.6-235850\\
89 & 270.103 & -23.979 & $20.61 \pm 4.69$ & $16.57 \pm 4.12$ & - & - & -6.15 & -14.71 & 29.97 & - & CXOU J180024.7-235845\\
90 & 270.136 & -23.973 & $20.12 \pm 5.39$ & $18.02 \pm 4.58$ & - & - & -5.39 & -13.3 & 31.38 & - & CXOU J180032.6-235822\\
91 & 270.039 & -23.966 & $64.76 \pm 9.75$ & $43.11 \pm 6.86$ & $18.46 \pm 4.9$ & - & -4.26 & -13.15 & 31.53 & - & CXOU J180009.3-235758\\
92 & 270.149 & -23.964 & $14.25 \pm 4.58$ & $12.2 \pm 3.74$ & - & - & -4.76 & -13.17 & 31.51 & - & CXOU J180035.7-235750\\
93 & 270.06 & -23.962 & $13.0 \pm 4.47$ & $11.22 \pm 3.61$ & - & - & -5.55 & -13.55 & 31.13 & - & CXOU J180014.4-235745\\
94 & 270.127 & -23.938 & $42.78 \pm 8.19$ & $38.66 \pm 6.56$ & - & - & -4.43 & -12.76 & 31.92 & - & CXOU J180030.4-235618\\
95 & 269.966 & -23.914 & $219.52 \pm 20.49$ & $49.69 \pm 10.25$ & $138.5 \pm 13.42$ & - & -2.44 & -11.47 & 33.21 & - & CXOU J175951.8-235452\\
96 & 270.163 & -24.156 & $44.53 \pm 9.64$ & $29.36 \pm 6.63$ & - & - & -4.85 & -12.75 & 31.93 & - & CXOU J180039.0-240921\\
97 & 270.084 & -24.138 & $37.97 \pm 8.72$ & $24.86 \pm 5.48$ & - & - & -4.81 & -12.7 & 31.98 & - & CXOU J180020.1-240816\\
98 & 270.129 & -24.133 & $15.98 \pm 4.9$ & $10.33 \pm 3.61$ & - & - & -5.79 & -13.84 & 30.84 & - & CXOU J180031.0-240800\\
99 & 270.151 & -24.127 & $17.53 \pm 5.2$ & $12.54 \pm 4.0$ & - & - & -5.94 & -14.15 & 30.53 & - & CXOU J180036.1-240738\\
100 & 270.067 & -24.123 & $26.14 \pm 6.25$ & $26.66 \pm 5.66$ & - & - & -5.13 & -13.03 & 31.65 & - & CXOU J180016.1-240723\\
\\
\hline\end{tabular}
\\
\begin{flushleft} 
  $\rm{^{A}}$ - Source identified in the conventional HII region, G5.89-0.39A\\
  $\rm{^{B}}$ - Source identified in the UC HII region, G5.89-0.39B\\
  $\rm{^{\ddagger}}$ - X-ray fluxes: Calculated for the full energy band (0.3-10.0 keV); P - photon flux; E - energy flux. 
  Average error in fluxes $\sim 25 \%$. Errors calculated from error in net counts.\\
  $\rm{^{\dagger}}$ - X-ray luminosities: f = full band (0.3-10.0 keV). Absorption corrected luminosity subscripted with a c. Luminosities were 
  calculated assuming a distance of 2 kpc. Average error in the luminosity is $\sim 25 \%$. Errors calculated from error in net counts.
\end{flushleft}
\end{table}
\end{landscape}

\begin{landscape}
\begin{table}
\caption{Table of Identified Sources and Their Values}
\setlength{\tabcolsep}{0pt}
\centering
\begin{tabular}{cccccccccccc}
\hline
 \multicolumn{1}{c}{} &
  \multicolumn{1}{c}{} &
  \multicolumn{1}{c}{} &
  \multicolumn{4}{c}{} &
  \multicolumn{2}{c}{Fluxes$^{\ddagger}$}&
  \multicolumn{2}{c}{Luminosities$^{\dagger}$} &
  \multicolumn{1}{c}{}
   \\
  
  \multicolumn{1}{c}{Source \#} &
  \multicolumn{1}{c}{RA} &
  \multicolumn{1}{c}{Dec} &
  \multicolumn{4}{c}{Net Counts} &
  \multicolumn{1}{c}{$\rm{log \, F_{P}}$}&
  \multicolumn{1}{c}{$\rm{log \, F_{E}}$} &
  \multicolumn{1}{c}{$\rm{log\, L_{f}}$} &
  \multicolumn{1}{c}{$\rm{log \, L_{f,c}}$} &
  \multicolumn{1}{c}{Chandra ID} \\
  &
   \multicolumn{1}{c}{degrees} & 
   \multicolumn{1}{c}{degrees} &
  \multicolumn{1}{c}{0.3-10 keV} &
  \multicolumn{1}{c}{0.3-2.5 keV} &
  \multicolumn{1}{c}{2.5-5 keV} &
  \multicolumn{1}{c}{5-10 keV} &
  \multicolumn{1}{c}{($\rm{ph\,cm^{-2}\,s^{-1}}$)}&
  \multicolumn{1}{c}{($\rm{erg\,cm^{-2}\,s^{-1}}$)}&
  \multicolumn{1}{c}{($\rm{erg\,s^{-1}}$)}&
  \multicolumn{1}{c}{($\rm{erg\,s^{-1}}$)}&
  \multicolumn{1}{c}{}
  \\
  \multicolumn{1}{c}{(1)}&
  \multicolumn{1}{c}{(2)}&
  \multicolumn{1}{c}{(3)}&
  \multicolumn{1}{c}{(4)}&
  \multicolumn{1}{c}{(5)}&
  \multicolumn{1}{c}{(6)}&
  \multicolumn{1}{c}{(7)}&
  \multicolumn{1}{c}{(8)}&
  \multicolumn{1}{c}{(9)}&
  \multicolumn{1}{c}{(10)}&
  \multicolumn{1}{c}{(11)} &
  \multicolumn{1}{c}{(12)}\\
\hline \\

101 & 270.187 & -24.108 & $22.25 \pm 5.48$ & $15.85 \pm 4.36$ & - & - & -5.33 & -13.22 & 31.46 & - & CXOU J180044.8-240629\\
102 & 270.178 & -24.108 & $32.37 \pm 7.07$ & $11.15 \pm 3.87$ & $13.76 \pm 4.12$ & - & -5.26 & -13.22 & 31.46 & - & CXOU J180042.8-240627\\
103 & 270.166 & -24.103 & $21.94 \pm 5.92$ & $8.17 \pm 3.32$ & $8.2 \pm 3.16$ & - & -5.04 & -12.97 & 31.71 & - & CXOU J180039.9-240611\\
104 & 270.084 & -24.099 & $14.32 \pm 4.24$ & - & $6.53 \pm 2.65$ & - & -5.71 & -13.65 & 31.03 & - & CXOU J180020.1-240556\\
105 & 270.197 & -24.081 & $20.97 \pm 5.57$ & $15.86 \pm 4.36$ & - & - & -5.33 & -13.34 & 31.34 & - & CXOU J180047.3-240452\\
106$\rm{^{A}}$ & 270.18 & -24.081 & $28.64 \pm 6.71$ & - & $9.62 \pm 3.46$ & - & -5.69 & -13.7 & 30.98 & 31.3 & CXOU J180043.3-240450\\
107 & 270.209 & -24.075 & $55.94 \pm 8.49$ & $54.83 \pm 8.0$ & - & - & -5.2 & -13.26 & 31.42 & - & CXOU J180050.2-240431\\
108$\rm{^{B}}$ & 270.135 & -24.075 & $13.17 \pm 3.87$ & - & $10.59 \pm 3.32$ & - & -5.79 & -13.68 & 31.0 & 31.32 & CXOU J180032.3-240431\\
109$\rm{^{A}}$ & 270.173 & -24.073 & $23.76 \pm 5.66$ & - & $13.77 \pm 4.12$ & - & -5.45 & -13.41 & 31.27 & 31.6 & CXOU J180041.4-240422\\
110$\rm{^{B}}$ & 270.128 & -24.073 & $10.69 \pm 3.46$ & $5.68 \pm 2.45$ & - & - & -6.29 & -14.45 & 30.23 & 30.55 & CXOU J180030.7-240423\\
111 & 270.036 & -24.069 & $15.84 \pm 4.9$ & $13.87 \pm 4.12$ & - & - & -5.73 & -13.99 & 30.69 & - & CXOU J180008.6-240407\\
112 & 270.221 & -24.065 & $11.93 \pm 6.0$ & $9.81 \pm 3.61$ & - & - & -4.83 & -12.85 & 31.83 & - & CXOU J180053.0-240353\\
113 & 270.193 & -24.056 & $16.89 \pm 5.2$ & $11.19 \pm 3.87$ & - & - & -5.05 & -13.65 & 31.03 & - & CXOU J180046.3-240321\\
114 & 269.995 & -24.054 & $32.43 \pm 8.43$ & - & $16.31 \pm 4.9$ & - & -5.63 & -13.76 & 30.92 & - & CXOU J175958.9-240315\\
115 & 270.058 & -24.052 & $13.44 \pm 4.12$ & - & $4.58 \pm 2.24$ & - & -6.0 & -14.09 & 30.59 & - & CXOU J180014.0-240308\\
116 & 270.192 & -24.039 & $14.24 \pm 4.8$ & $9.25 \pm 3.46$ & - & - & -5.14 & -13.43 & 31.25 & - & CXOU J180046.1-240221\\
117 & 270.035 & -24.036 & $16.77 \pm 5.1$ & $9.19 \pm 3.32$ & - & - & -5.28 & -13.15 & 31.53 & - & CXOU J180008.3-240209\\
118 & 270.25 & -24.028 & $29.03 \pm 8.06$ & $28.36 \pm 6.4$ & - & - & -4.6 & -12.57 & 32.11 & - & CXOU J180100.1-240142\\
119 & 270.001 & -24.024 & $33.94 \pm 8.12$ & $33.4 \pm 6.86$ & - & - & -4.24 & -12.74 & 31.94 & - & CXOU J180000.2-240125\\
120 & 270.056 & -24.021 & $14.28 \pm 4.24$ & - & $10.26 \pm 3.32$ & - & -6.38 & -14.68 & 30.0 & - & CXOU J180013.4-240117\\

\\
\hline\end{tabular}
\\
\begin{flushleft} 
  $\rm{^{A}}$ - Source identified in the conventional HII region, G5.89-0.39A\\
  $\rm{^{B}}$ - Source identified in the UC HII region, G5.89-0.39B\\
  $\rm{^{\ddagger}}$ - X-ray fluxes: Calculated for the full energy band (0.3-10.0 keV); P - photon flux; E - energy flux. 
  Average error in fluxes $\sim 25 \%$. Errors calculated from error in net counts.\\
  $\rm{^{\dagger}}$ - X-ray luminosities: f = full band (0.3-10.0 keV). Absorption corrected luminosity subscripted with a c. Luminosities were 
  calculated assuming a distance of 2 kpc. Average error in the luminosity is $\sim 25 \%$. Errors calculated from error in net counts.
\end{flushleft}
\end{table}
\end{landscape}

\begin{landscape}
\begin{table}
\caption{Table of Identified Sources and Their Values}
\setlength{\tabcolsep}{0pt}
\centering
\begin{tabular}{cccccccccccc}
\hline
 \multicolumn{1}{c}{} &
  \multicolumn{1}{c}{} &
  \multicolumn{1}{c}{} &
  \multicolumn{4}{c}{} &
  \multicolumn{2}{c}{Fluxes$^{\ddagger}$}&
  \multicolumn{2}{c}{Luminosities$^{\dagger}$} &
  \multicolumn{1}{c}{}
   \\
  
  \multicolumn{1}{c}{Source \#} &
  \multicolumn{1}{c}{RA} &
  \multicolumn{1}{c}{Dec} &
  \multicolumn{4}{c}{Net Counts} &
  \multicolumn{1}{c}{$\rm{log \, F_{P}}$}&
  \multicolumn{1}{c}{$\rm{log \, F_{E}}$} &
  \multicolumn{1}{c}{$\rm{log\, L_{f}}$} &
  \multicolumn{1}{c}{$\rm{log \, L_{f,c}}$} &
  \multicolumn{1}{c}{Chandra ID} \\
  &
   \multicolumn{1}{c}{degrees} & 
   \multicolumn{1}{c}{degrees} &
  \multicolumn{1}{c}{0.3-10 keV} &
  \multicolumn{1}{c}{0.3-2.5 keV} &
  \multicolumn{1}{c}{2.5-5 keV} &
  \multicolumn{1}{c}{5-10 keV} &
  \multicolumn{1}{c}{($\rm{ph\,cm^{-2}\,s^{-1}}$)}&
  \multicolumn{1}{c}{($\rm{erg\,cm^{-2}\,s^{-1}}$)}&
  \multicolumn{1}{c}{($\rm{erg\,s^{-1}}$)}&
  \multicolumn{1}{c}{($\rm{erg\,s^{-1}}$)}&
  \multicolumn{1}{c}{}
  \\
  \multicolumn{1}{c}{(1)}&
  \multicolumn{1}{c}{(2)}&
  \multicolumn{1}{c}{(3)}&
  \multicolumn{1}{c}{(4)}&
  \multicolumn{1}{c}{(5)}&
  \multicolumn{1}{c}{(6)}&
  \multicolumn{1}{c}{(7)}&
  \multicolumn{1}{c}{(8)}&
  \multicolumn{1}{c}{(9)}&
  \multicolumn{1}{c}{(10)}&
  \multicolumn{1}{c}{(11)} &
  \multicolumn{1}{c}{(12)}\\
\hline \\
121 & 270.037 & -24.003 & $10.98 \pm 4.0$ & - & $10.18 \pm 3.46$ & - & -5.48 & -13.33 & 31.35 & - & CXOU J180008.8-240011\\
122 & 270.213 & -24.002 & $24.45 \pm 7.35$ & $19.88 \pm 4.9$ & - & - & -4.91 & -12.9 & 31.78 & - & CXOU J180051.2-240008\\
123 & 270.206 & -23.995 & $24.34 \pm 6.0$ & $19.79 \pm 4.9$ & - & - & -5.1 & -13.55 & 31.13 & - & CXOU J180049.5-235942\\
124 & 270.154 & -23.99 & $17.29 \pm 4.58$ & - & $13.36 \pm 3.74$ & - & -5.89 & -13.92 & 30.76 & - & CXOU J180036.9-235924\\
125 & 270.034 & -23.989 & $14.52 \pm 4.8$ & $7.72 \pm 3.16$ & - & - & -5.21 & -13.22 & 31.46 & - & CXOU J180008.1-235922\\
126 & 270.038 & -23.977 & $20.93 \pm 5.39$ & $21.81 \pm 5.0$ & - & - & -5.75 & -13.85 & 30.83 & - & CXOU J180009.1-235837\\
127 & 270.063 & -23.968 & $16.19 \pm 4.8$ & $7.54 \pm 3.0$ & - & - & -5.16 & -13.07 & 31.61 & - & CXOU J180015.2-235805\\
128 & 270.141 & -23.961 & $20.2 \pm 5.29$ & $8.39 \pm 3.16$ & $8.18 \pm 3.16$ & - & -5.31 & -13.33 & 31.35 & - & CXOU J180033.8-235739\\
129 & 270.079 & -23.947 & $18.94 \pm 6.4$ & $16.9 \pm 4.47$ & - & - & -5.1 & -12.98 & 31.7 & - & CXOU J180019.0-235650\\
130 & 270.066 & -23.941 & $52.17 \pm 9.11$ & $52.96 \pm 8.43$ & - & - & -5.07 & -13.09 & 31.59 & - & CXOU J180015.8-235626\\
131 & 269.996 & -23.933 & $130.02 \pm 16.34$ & $13.09 \pm 4.58$ & $82.1 \pm 10.86$ & $29.56 \pm 8.19$ & -3.32 & -11.9 & 32.78 & - & CXOU J175959.1-235560\\
132 & 270.154 & -23.932 & $28.91 \pm 7.42$ & - & $20.94 \pm 5.29$ & - & -4.69 & -12.83 & 31.85 & - & CXOU J180037.0-235555\\
133 & 270.011 & -23.92 & $139.37 \pm 19.52$ & $123.85 \pm 15.23$ & - & - & -2.11 & -11.36 & 33.32 & - & CXOU J180002.7-235513\\
134 & 270.069 & -23.913 & $65.0 \pm 10.68$ & - & $43.18 \pm 7.48$ & - & -4.24 & -12.55 & 32.13 & - & CXOU J180016.5-235446\\
135 & 270.024 & -24.181 & $70.78 \pm 15.2$ & $59.06 \pm 11.36$ & - & - & -3.61 & -12.26 & 32.42 & - & CXOU J180005.7-241052\\
136 & 270.078 & -24.171 & $26.2 \pm 7.94$ & $27.88 \pm 6.71$ & - & - & -4.69 & -13.34 & 31.34 & - & CXOU J180018.8-241014\\
137 & 270.184 & -24.143 & $22.08 \pm 7.21$ & - & $25.81 \pm 5.83$ & - & -4.66 & -12.79 & 31.89 & - & CXOU J180044.1-240836\\
138 & 270.067 & -24.135 & $45.76 \pm 9.7$ & $17.74 \pm 4.69$ & - & - & -3.63 & -12.27 & 32.41 & - & CXOU J180016.1-240806\\
139 & 270.114 & -24.127 & $24.68 \pm 6.25$ & $24.07 \pm 5.39$ & - & - & -5.66 & -14.16 & 30.52 & - & CXOU J180027.3-240738\\
140 & 270.049 & -24.123 & $27.83 \pm 7.87$ & - & $11.82 \pm 3.74$ & - & -3.47 & -12.41 & 32.27 & - & CXOU J180011.7-240722\\

\\
\hline\end{tabular}
\\
\begin{flushleft} 
  $\rm{^{A}}$ - Source identified in the conventional HII region, G5.89-0.39A\\
  $\rm{^{B}}$ - Source identified in the UC HII region, G5.89-0.39B\\
  $\rm{^{\ddagger}}$ - X-ray fluxes: Calculated for the full energy band (0.3-10.0 keV); P - photon flux; E - energy flux. 
  Average error in fluxes $\sim 25 \%$. Errors calculated from error in net counts.\\
  $\rm{^{\dagger}}$ - X-ray luminosities: f = full band (0.3-10.0 keV). Absorption corrected luminosity subscripted with a c. Luminosities were 
  calculated assuming a distance of 2 kpc. Average error in the luminosity is $\sim 25 \%$. Errors calculated from error in net counts.
\end{flushleft}
\end{table}
\end{landscape}

\begin{landscape}
\begin{table}
\caption{Table of Identified Sources and Their Values}
\setlength{\tabcolsep}{0pt}
\centering
\begin{tabular}{cccccccccccc}
\hline
 \multicolumn{1}{c}{} &
  \multicolumn{1}{c}{} &
  \multicolumn{1}{c}{} &
  \multicolumn{4}{c}{} &
  \multicolumn{2}{c}{Fluxes$^{\ddagger}$}&
  \multicolumn{2}{c}{Luminosities$^{\dagger}$} &
  \multicolumn{1}{c}{}
   \\
  
  \multicolumn{1}{c}{Source \#} &
  \multicolumn{1}{c}{RA} &
  \multicolumn{1}{c}{Dec} &
  \multicolumn{4}{c}{Net Counts} &
  \multicolumn{1}{c}{$\rm{log \, F_{P}}$}&
  \multicolumn{1}{c}{$\rm{log \, F_{E}}$} &
  \multicolumn{1}{c}{$\rm{log\, L_{f}}$} &
  \multicolumn{1}{c}{$\rm{log \, L_{f,c}}$} &
  \multicolumn{1}{c}{Chandra ID}\\
  &
   \multicolumn{1}{c}{degrees} & 
   \multicolumn{1}{c}{degrees} &
  \multicolumn{1}{c}{0.3-10 keV} &
  \multicolumn{1}{c}{0.3-2.5 keV} &
  \multicolumn{1}{c}{2.5-5 keV} &
  \multicolumn{1}{c}{5-10 keV} &
  \multicolumn{1}{c}{($\rm{ph\,cm^{-2}\,s^{-1}}$)}&
  \multicolumn{1}{c}{($\rm{erg\,cm^{-2}\,s^{-1}}$)}&
  \multicolumn{1}{c}{($\rm{erg\,s^{-1}}$)}&
  \multicolumn{1}{c}{($\rm{erg\,s^{-1}}$)} &
  \multicolumn{1}{c}{}
  \\
  \multicolumn{1}{c}{(1)}&
  \multicolumn{1}{c}{(2)}&
  \multicolumn{1}{c}{(3)}&
  \multicolumn{1}{c}{(4)}&
  \multicolumn{1}{c}{(5)}&
  \multicolumn{1}{c}{(6)}&
  \multicolumn{1}{c}{(7)}&
  \multicolumn{1}{c}{(8)}&
  \multicolumn{1}{c}{(9)}&
  \multicolumn{1}{c}{(10)}&
  \multicolumn{1}{c}{(11)}&
  \multicolumn{1}{c}{(12)}\\
\hline \\
141 & 270.234 & -24.113 & $42.89 \pm 9.8$ & $15.51 \pm 5.2$ & - & - & -3.94 & -12.25 & 32.43 & - & CXOU J180056.1-240648\\
142 & 269.971 & -24.105 & $39.05 \pm 9.11$ & $22.86 \pm 6.16$ & - & - & -4.22 & -12.44 & 32.24 & - & CXOU J175953.0-240618\\
143 & 269.974 & -24.091 & $22.37 \pm 7.55$ & $15.45 \pm 5.0$ & - & - & -4.96 & -12.86 & 31.82 & - & CXOU J175953.8-240529\\
144 & 269.984 & -24.066 & $21.09 \pm 6.86$ & $16.16 \pm 4.9$ & - & - & -2.62 & -11.89 & 32.79 & - & CXOU J175956.1-240357\\
145 & 269.995 & -24.058 & $23.67 \pm 7.42$ & $18.92 \pm 5.57$ & - & - & -4.38 & -12.89 & 31.79 & - & CXOU J175958.7-240330\\
146 & 270.232 & -24.05 & $111.54 \pm 14.35$ & $48.02 \pm 8.72$ & $49.24 \pm 8.37$ & - & -4.34 & -12.53 & 32.15 & - & CXOU J180055.7-240260\\
147 & 270.358 & -24.02 & $317.16 \pm 35.07$ & - & $207.31 \pm 20.54$ & - & -2.97 & -10.87 & 33.81 & - & CXOU J180126.0-240110\\
148 & 270.184 & -23.964 & $36.09 \pm 9.0$ & - & $10.23 \pm 3.46$ & - & -4.53 & -13.09 & 31.59 & - & CXOU J180044.3-235751\\
149 & 269.999 & -23.941 & $24.37 \pm 9.49$ & $31.77 \pm 7.21$ & - & - & -4.32 & -12.31 & 32.37 & - & CXOU J175959.7-235628\\
150 & 270.088 & -23.936 & $22.15 \pm 7.35$ & - & - & $12.15 \pm 4.58$ & -4.36 & -12.85 & 31.83 & - & CXOU J180021.1-235611\\
151 & 270.019 & -23.933 & $20.26 \pm 6.48$ & $20.92 \pm 5.48$ & - & - & -5.14 & -13.52 & 31.16 & - & CXOU J180004.5-235560\\
152 & 270.033 & -23.925 & $36.26 \pm 9.0$ & - & $24.21 \pm 5.66$ & - & -5.17 & -13.19 & 31.49 & - & CXOU J180007.9-235532\\
153 & 270.048 & -23.91 & $25.5 \pm 7.55$ & $21.96 \pm 6.16$ & - & - & -4.08 & -12.69 & 31.99 & - & CXOU J180011.5-235437\\
154 & 270.196 & -24.176 & $44.67 \pm 11.71$ & $20.62 \pm 5.83$ & - & - & -4.83 & -12.93 & 31.75 & - & CXOU J180047.1-241035\\
155 & 270.314 & -24.172 & $86.57 \pm 15.97$ & - & $44.03 \pm 10.3$ & - & -4.33 & -12.22 & 32.46 & - & CXOU J180115.3-241019\\
156 & 270.248 & -24.08 & $37.43 \pm 10.0$ & $32.72 \pm 7.87$ & - & - & -4.92 & -12.84 & 31.84 & - & CXOU J180059.6-240449\\
157 & 269.973 & -24.079 & $31.43 \pm 8.37$ & $15.75 \pm 5.0$ & - & - & -4.96 & -12.83 & 31.85 & - & CXOU J175953.5-240444\\
158 & 269.959 & -24.019 & $30.55 \pm 8.43$ & $20.95 \pm 5.83$ & - & - & -3.11 & -12.08 & 32.6 & - & CXOU J175950.1-240107\\
159 & 270.219 & -24.07 & - & $13.45 \pm 4.36$ & $8.31 \pm 3.32$ & -  & -5.36 & -13.33 & 31.35 & - & CXOU J180052.6-240413\\

\\
\hline\end{tabular}
\\
\begin{flushleft} 
  $\rm{^{A}}$ - Source identified in the conventional HII region, G5.89-0.39A\\
  $\rm{^{B}}$ - Source identified in the UC HII region, G5.89-0.39B\\
  $\rm{^{\ddagger}}$ - X-ray fluxes: Calculated for the full energy band (0.3-10.0 keV); P - photon flux; E - energy flux. 
  Average error in fluxes $\sim 25 \%$. Errors calculated from error in net counts.\\
  $\rm{^{\dagger}}$ - X-ray luminosities: f = full band (0.3-10.0 keV). Absorption corrected luminosity subscripted with a c. Luminosities were 
  calculated assuming a distance of 2 kpc. Average error in luminosity $\sim 25 \%$. Errors calculated from error in net counts.
\end{flushleft}
\label{tab:fin}
\end{table}
\end{landscape}

\begin{table}
\caption{Table of Identified Sources - Continued}
\setlength{\tabcolsep}{0pt}
\centering
\begin{tabular}{cccc}
\hline
 \multicolumn{1}{c}{} &
  \multicolumn{1}{c}{} &
  \multicolumn{1}{c}{} &
  \multicolumn{1}{c}{} 
   \\
  
  \multicolumn{1}{c}{Source \#} &
  \multicolumn{1}{c}{RA} &
  \multicolumn{1}{c}{Dec} &
  \multicolumn{1}{c}{Source Variability Index} \\
  &
   \multicolumn{1}{c}{degrees} & 
   \multicolumn{1}{c}{degrees} &
  \\
  \multicolumn{1}{c}{(1)}&
  \multicolumn{1}{c}{(2)}&
  \multicolumn{1}{c}{(3)}&
  \multicolumn{1}{c}{(13)}\\
\hline \\
1 & 270.08 & -24.134 & 0\\
2 & 270.142 & -24.098 & 6\\
3 & 270.144 & -24.077 & 5\\
4$\rm{^{B}}$ & 270.129 & -24.077 & 2\\
5$\rm{^{B}}$ & 270.144 & -24.074 & 7\\
6$\rm{^{B}}$ & 270.128 & -24.071 & 0\\
7$\rm{^{B}}$ & 270.124 & -24.069 & 0\\
8$\rm{^{A}}$ & 270.171 & -24.069 & 0\\
9$\rm{^{B}}$ & 270.131 & -24.068 & 6\\
10$\rm{^{B}}$ & 270.129 & -24.068 & 1\\
11 & 270.089 & -24.067 & 1\\
12$\rm{^{B}}$ & 270.118 & -24.066 & 1\\
13$\rm{^{B}}$ & 270.127 & -24.062 & 2\\
14 & 270.115 & -24.059 &  0\\
15$\rm{^{A}}$ & 270.168 & -24.052 & 6\\
16 & 270.136 & -24.052 & 9\\
17 & 270.121 & -24.05 & 1\\
18 & 270.143 & -24.049 & 6\\
19 & 270.156 & -24.048 & 1\\
20 & 270.056 & -24.032 & 1\\
21 & 270.18 & -24.03 & 2\\
22 & 270.128 & -24.026 & 8\\
23 & 270.042 & -24.025 & 0\\
24 & 270.126 & -24.023 & 0\\
25 & 270.177 & -24.022 & 0\\
26 & 270.141 & -24.015 & 0\\
27 & 270.07 & -24.008 & 2\\
28 & 270.135 & -24.008 & 0\\
29 & 270.13 & -24.002 & 0\\
30 & 270.133 & -23.994 & 7\\
31 & 270.091 & -24.142 & 2\\
32 & 270.072 & -24.13 & 2\\
33 & 270.136 & -24.102 & 0\\
34 & 270.106 & -24.1 & 0\\
35 & 270.123 & -24.096 & 0\\
36$\rm{^{A}}$ & 270.164 & -24.092 & 1\\
37 & 270.129 & -24.092 & 2\\
38 & 270.083 & -24.09 & 1\\
39 & 270.2 & -24.089 & 7\\
40 & 270.152 & -24.088 & 1\\
\\
\hline\end{tabular}
\\
\begin{flushleft} 
  $\rm{^{A}}$ - Source identified in the conventional HII region, G5.89-0.39A\\
  $\rm{^{B}}$ - Source identified in the UC HII region, G5.89-0.39B\\
\end{flushleft}
\label{tab:vars1}
\end{table}

\begin{table}
\caption{Table of Identified Sources - Continued}
\setlength{\tabcolsep}{0pt}
\centering
\begin{tabular}{cccc}
\hline
 \multicolumn{1}{c}{} &
  \multicolumn{1}{c}{} &
  \multicolumn{1}{c}{} &
  \multicolumn{1}{c}{} 
   \\
  
  \multicolumn{1}{c}{Source \#} &
  \multicolumn{1}{c}{RA} &
  \multicolumn{1}{c}{Dec} &
  \multicolumn{1}{c}{Source Variability Index} \\
  &
   \multicolumn{1}{c}{degrees} & 
   \multicolumn{1}{c}{degrees} &
  \\
  \multicolumn{1}{c}{(1)}&
  \multicolumn{1}{c}{(2)}&
  \multicolumn{1}{c}{(3)}&
  \multicolumn{1}{c}{(13)}\\
\hline \\
41 & 270.115 & -24.088 & 0\\
42 & 270.108 & -24.085 & 9\\
43$\rm{^{A}}$ & 270.164 & -24.085 & 6\\
44$\rm{^{A}}$ & 270.178 & -24.085 & 7\\
45 & 270.117 & -24.082 & 0\\
46$\rm{^{A}}$ & 270.17 & -24.082 & 0\\
47 & 270.142 & -24.079 & 0\\
48 & 270.15 & -24.079 & 7\\
49$\rm{^{B}}$ & 270.125 & -24.077 & 7\\
50$\rm{^{B}}$ & 270.137 & -24.077 & 2\\
51$\rm{^{A}}$ & 270.175 & -24.077 & 0\\
52 & 270.049 & -24.075 & 1\\
53$\rm{^{A}}$ & 270.176 & -24.074 & 0\\
54 & 270.075 & -24.073 & 2\\
55$\rm{^{A}}$ & 270.177 & -24.072 & 0\\
56$\rm{^{A}}$ & 270.179 & -24.072 & 8\\
57$\rm{^{A}}$ & 270.161 & -24.071 & 2\\
58$\rm{^{A}}$ & 270.17 & -24.071 & 0\\
59$\rm{^{A}}$ & 270.176 & -24.071 & 0\\
60$\rm{^{A}}$ & 270.164 & -24.07 & 2\\
61$\rm{^{A}}$ & 270.181 & -24.068 & 0\\
62$\rm{^{B}}$ & 270.126 & -24.067 & 0\\
63 & 270.05 & -24.066 & 0\\
64 & 270.076 & -24.065 & 2\\
65$\rm{^{A}}$ & 270.155 & -24.065 & 1\\
66$\rm{^{B}}$ & 270.133 & -24.064 & 2\\
67 & 270.084 & -24.064 & 2\\
68$\rm{^{A}}$ & 270.167 & -24.064 & 2\\
69$\rm{^{A}}$ & 270.18 & -24.064 & 0\\
70$\rm{^{B}}$ & 270.131 & -24.063 & 0\\
71 & 270.053 & -24.059 & 7\\
72 & 270.196 & -24.056 & 6\\
73$\rm{^{A}}$ & 270.177 & -24.054 & 6\\
74$\rm{^{A}}$ & 270.182 & -24.054 & 0\\
75 & 270.033 & -24.051 & 6\\
76 & 270.176 & -24.047 & 0\\
77 & 270.174 & -24.032 & 2\\
78 & 270.061 & -24.032 & 1\\
79 & 270.06 & -24.031 & 1\\
80 & 270.178 & -24.03 & 0\\
\\
\hline\end{tabular}
\\
\begin{flushleft} 
  $\rm{^{A}}$ - Source identified in the conventional HII region, G5.89-0.39A\\
  $\rm{^{B}}$ - Source identified in the UC HII region, G5.89-0.39B\\
\end{flushleft}
\label{tab:vars2}
\end{table}

\begin{table}
\caption{Table of Identified Sources - Continued}
\setlength{\tabcolsep}{0pt}
\centering
\begin{tabular}{cccc}
\hline
 \multicolumn{1}{c}{} &
  \multicolumn{1}{c}{} &
  \multicolumn{1}{c}{} &
  \multicolumn{1}{c}{} 
   \\
  
  \multicolumn{1}{c}{Source \#} &
  \multicolumn{1}{c}{RA} &
  \multicolumn{1}{c}{Dec} &
  \multicolumn{1}{c}{Source Variability Index} \\
  &
   \multicolumn{1}{c}{degrees} & 
   \multicolumn{1}{c}{degrees} &
  \\
  \multicolumn{1}{c}{(1)}&
  \multicolumn{1}{c}{(2)}&
  \multicolumn{1}{c}{(3)}&
  \multicolumn{1}{c}{(13)}\\
\hline \\
81 & 270.068 & -24.028 & 0\\
82 & 270.065 & -24.025 & 5\\
83 & 270.062 & -24.023 & 0\\
84 & 270.027 & -24.022 & 8\\
85 & 270.179 & -24.019 & 2\\
86 & 270.035 & -23.999 & 1\\
87 & 270.048 & -23.99 & 0\\
88 & 270.073 & -23.981 & 0\\
89 & 270.103 & -23.979 & 0\\
90 & 270.136 & -23.973 & 0\\
91 & 270.039 & -23.966 & 0\\
92 & 270.149 & -23.964 & 0\\
93 & 270.06 & -23.962 & 0\\
94 & 270.127 & -23.938 & 0\\
95 & 269.966 & -23.914 & 0\\
96 & 270.163 & -24.156 & 0\\
97 & 270.084 & -24.138 & 0\\
98 & 270.129 & -24.133 & 0\\
99 & 270.151 & -24.127 & 6\\
100 & 270.067 & -24.123 & 0\\
101 & 270.187 & -24.108 & 0\\
102 & 270.178 & -24.108 & 7\\
103 & 270.166 & -24.103 & 0\\
104 & 270.084 & -24.099 & 0\\
105 & 270.197 & -24.081 & \\
106$\rm{^{A}}$ & 270.18 & -24.081 & 0\\
107 & 270.209 & -24.075 & 0\\
108$\rm{^{B}}$ & 270.135 & -24.075 & 0\\
109$\rm{^{A}}$ & 270.173 & -24.073 & 0\\
110$\rm{^{B}}$ & 270.128 & -24.073 & 2\\
111 & 270.036 & -24.069 & 2\\
112 & 270.221 & -24.065 & 0\\
113 & 270.193 & -24.056 & 1\\
114 & 269.995 & -24.054 & 0\\
115 & 270.058 & -24.052 & 2\\
116 & 270.192 & -24.039 & 0\\
117 & 270.035 & -24.036 & 0\\
118 & 270.25 & -24.028 & 0\\
119 & 270.001 & -24.024 & 1\\
120 & 270.056 & -24.021 & 0\\
\\
\hline\end{tabular}
\\
\begin{flushleft} 
  $\rm{^{A}}$ - Source identified in the conventional HII region, G5.89-0.39A\\
  $\rm{^{B}}$ - Source identified in the UC HII region, G5.89-0.39B\\
\end{flushleft}
\label{tab:vars3}
\end{table}

\begin{table}
\caption{Table of Identified Sources - Continued}
\setlength{\tabcolsep}{0pt}
\centering
\begin{tabular}{cccc}
\hline
 \multicolumn{1}{c}{} &
  \multicolumn{1}{c}{} &
  \multicolumn{1}{c}{} &
  \multicolumn{1}{c}{} 
   \\
  
  \multicolumn{1}{c}{Source \#} &
  \multicolumn{1}{c}{RA} &
  \multicolumn{1}{c}{Dec} &
  \multicolumn{1}{c}{Source Variability Index} \\
  &
   \multicolumn{1}{c}{degrees} & 
   \multicolumn{1}{c}{degrees} &
  \\
  \multicolumn{1}{c}{(1)}&
  \multicolumn{1}{c}{(2)}&
  \multicolumn{1}{c}{(3)}&
  \multicolumn{1}{c}{(13)}\\
\hline \\
121 & 270.037 & -24.003 & 0\\
122 & 270.213 & -24.002 & 1\\
123 & 270.206 & -23.995 & 1\\
124 & 270.154 & -23.99 & 2\\
125 & 270.034 & -23.989 & 1\\
126 & 270.038 & -23.977 & 1\\
127 & 270.063 & -23.968 & 2\\
128 & 270.141 & -23.961 & 0\\
129 & 270.079 & -23.947 & 0\\
130 & 270.066 & -23.941 & 0\\
131 & 269.996 & -23.933 & 0\\
132 & 270.154 & -23.932 & 0\\
133 & 270.011 & -23.92 & 0\\
134 & 270.069 & -23.913 & 0\\
135 & 270.024 & -24.181 & 0\\
136 & 270.078 & -24.171 & 0\\
137 & 270.184 & -24.143 & 1\\
138 & 270.067 & -24.135 & 0\\
139 & 270.114 & -24.127 & 0\\
140 & 270.049 & -24.123 & 2\\
141 & 270.234 & -24.113 & 0\\
142 & 269.971 & -24.105 & 1\\
143 & 269.974 & -24.091 & 2\\
144 & 269.984 & -24.066 & 0\\
145 & 269.995 & -24.058 & 0\\
146 & 270.232 & -24.05 & 0\\
147 & 270.358 & -24.02 & 0\\
148 & 270.184 & -23.964 & 0\\
149 & 269.999 & -23.941 & 0\\
150 & 270.088 & -23.936 & 0\\
151 & 270.019 & -23.933 & 1\\
152 & 270.033 & -23.925 & 2\\
153 & 270.048 & -23.91 & 2\\
154 & 270.196 & -24.176 & 7\\
155 & 270.314 & -24.172 & 0\\
156 & 270.248 & -24.08 & 0\\
157 & 269.973 & -24.079 & 7\\
158 & 269.959 & -24.019 & 0\\
159 & 270.219 & -24.07 & 0\\

\\
\hline\end{tabular}
\\
\begin{flushleft} 
  $\rm{^{A}}$ - Source identified in the conventional HII region, G5.89-0.39A\\
  $\rm{^{B}}$ - Source identified in the UC HII region, G5.89-0.39B\\
\end{flushleft}
\label{tab:vars4}
\end{table}

\label{lastpage}

\end{document}